\documentclass[10pt,english,two column]{IEEEtran}
\usepackage[T1]{fontenc}
\usepackage[latin9]{inputenc}
\usepackage{float}
\usepackage{amsmath}
\usepackage{amsthm}
\usepackage{amssymb}
\usepackage{graphicx}
\usepackage{stfloats}
\usepackage{cite} 
\usepackage{xcolor}

\makeatletter

\floatstyle{ruled}
\newfloat{algorithm}{tbp}{loa}
\providecommand{\algorithmname}{Algorithm}
\floatname{algorithm}{\protect\algorithmname}

\theoremstyle{plain}

\theoremstyle{plain}

\ifCLASSINFOpdf
\else
\fi
\usepackage{babel}

\makeatother

\usepackage{babel}
\providecommand{\propositionname}{Proposition}
\providecommand{\theoremname}{Theorem}

\begin{document}

\title{Collaborative Cloud and Edge Mobile Computing in C-RAN Systems with Minimal End-to-End Latency}

\author{Seok-Hwan Park, \textit{Member}, \textit{IEEE}, Seongah Jeong, \textit{Member}, \textit{IEEE}, Jinyeop Na, \textit{Student Member}, \textit{IEEE}, \\ Osvaldo Simeone, \textit{Fellow}, \textit{IEEE}, and Shlomo Shamai (Shitz), \textit{Life Fellow}, \textit{IEEE} \thanks{S.-H. Park was supported by Basic Science Research Program through the National Research Foundation of Korea (NRF) grants funded by the Ministry of Education [NRF-2019R1A6A1A09031717, 2021R1C1C1006557]. The work of S. Jeong was 
supported by the MSIT (Ministry of Science and ICT), Korea, under the ITRC (Information Technology Research Center) support program (IITP-2020-0-01787) supervised by the IITP (Institute of Information \& Communications Technology Planning \& Evaluation). This work was also supported by the European Research Council (ERC) under the European Union's Horizon 2020
Research and Innovation Programme (Grant Agreement Nos. 694630 and 725731).

S.-H. Park is with the Division of Electronic Engineering and the Future Semiconductor Convergence Technology Research Center, Jeonbuk
National University, Jeonju 54896, Korea (email: seokhwan@jbnu.ac.kr).

S. Jeong is with the School of Electronics Engineering, Kyungpook National University, Daegu 14566, Korea (email: seongah@knu.ac.kr).

J. Na is with the Department of Electrical Engineering, Korea Advanced Institute of Science and Technology (KAIST), Daejeon 34141, Korea (email: wlsduq37@kaist.ac.kr).

O. Simeone is with King's Communication, Learning and Information Processing (kclip) Lab, the Centre for Telecommunications Research, Department of Engineering, King's College London, London WC2R 2LS, U.K (email: osvaldo.simeone@kcl.ac.uk).

S. Shamai is with the Department of Electrical and Computer Engineering, Technion, Haifa 3200003, Israel (email: sshlomo@ee.technion.ac.il).
}}
\maketitle
\begin{abstract}
Mobile cloud and edge computing protocols make it possible to offer computationally heavy applications to mobile devices via computational offloading from devices to nearby edge servers or more powerful, but remote, cloud servers. Previous work assumed that computational tasks can be fractionally offloaded at both cloud processor (CP) and at a local edge node (EN) within a conventional Distributed Radio Access Network (D-RAN) that relies on  non-cooperative ENs equipped with one-way uplink fronthaul connection to the cloud.
In this paper, we propose to integrate collaborative fractional computing across CP and ENs within a Cloud RAN (C-RAN) architecture with finite-capacity two-way fronthaul links. Accordingly, tasks offloaded by a mobile device can be partially carried out at an EN and the CP, with multiple ENs communicating with a common CP to exchange data and computational outcomes while allowing for centralized precoding and decoding. Unlike prior work, we investigate joint optimization of computing and communication resources, including wireless and fronthaul segments, to minimize the end-to-end latency by accounting for a two-way uplink and downlink transmission. The problem is tackled by using fractional programming (FP) and matrix FP. Extensive numerical results validate the performance gain of the proposed architecture as compared to the previously studied D-RAN solution.
\end{abstract}


\theoremstyle{theorem}
\newtheorem{theorem}{Theorem}
\theoremstyle{proposition}
\newtheorem{proposition}{Proposition}
\theoremstyle{lemma}
\newtheorem{lemma}{Lemma}
\theoremstyle{corollary}
\newtheorem{corollary}{Corollary}
\theoremstyle{definition}
\newtheorem{definition}{Definition}
\theoremstyle{remark}
\newtheorem{remark}{Remark}

\section{Introduction} \label{sec:intro}

Mobile cloud and edge computing techniques enable computationally heavy applications such as gaming and augmented reality (AR) by offloading computation tasks from battery-limited mobile user equipments (UEs) to cloud or edge servers which are located respectively at cloud processor (CP) or edge nodes (ENs) of a cellular architecture \cite{Dinh-et-al:WCMC13, Satyanarayanan-et-al:PC09, Sardellitti:TSIPN15, Ashuwaili:TSIPN17, Tran-et-al:CM17, Mach-Becvar:CST17, Xiao-et-al:FGCS20}. In systems with both cloud and edge computing capabilities, computation tasks can be opportunistically offloaded either to ENs or to the CP \cite{Ren-et-al:TVT19}. For example, it may be desirable to offload latency-insensitive and computationally heavy tasks to a CP, while relatively light tasks with more stringent latency constraints can be offloaded to edge servers in ENs.

The optimization of the offloading decision policy was studied in \cite{Kumar-Lu:Computer10, Huang-et-al:TWC12} by focusing on the application layer and without including constraints imposed by the Radio Access Network (RAN).
To the best of our knowledge, reference \cite{Sardellitti:TSIPN15} for the first time studied  the \textit{joint} optimization of computation and communication resources for mobile wireless edge computing systems, with follow-up works including \cite{Ashuwaili:TSIPN17}. Both papers \cite{Sardellitti:TSIPN15, Ashuwaili:TSIPN17} aimed at minimizing energy expenditure under constraints on the end-to-end latency that encompass the contributions of both communication and computation. 
While \cite{Sardellitti:TSIPN15} accounts only for uplink transmission, reference \cite{Ashuwaili:TSIPN17} also includes the contribution of downlink communication, which is required to feed back the results of the remote computations.
To overcome the inherent non-convexity of the resulting optimization problems, the authors in \cite{Sardellitti:TSIPN15, Ashuwaili:TSIPN17} applied successive convex approximation (SCA) \cite{Scutari-et-al:TSP17:Part1, Scutari-et-al:TSP17:Part2}, which efficiently finds a locally optimal solution for constrained non-convex problems.
Extensions in \cite{Ashuwaili:WCL17, Jeong:TVT18} studied edge computing-based AR applications \cite{Ashuwaili:WCL17} and edge computing via an unmanned aerial vehicle (UAV) mounted cloudlet \cite{Jeong:TVT18}.

In a system with both \textit{cloud and edge computing} capabilities, computation tasks can be partially offloaded to CP and ENs \cite{Ren-et-al:TVT19}. Reference \cite{Ren-et-al:TVT19} tackled the problem of jointly optimizing communication and computational resources with the goal of minimizing a weighted sum of per-UE end-to-end latency metrics
within a distributed RAN (D-RAN) architecture \cite[Sec. III]{Kang-et-al:TWC18}. The authors in \cite{Ren-et-al:TVT19} developed closed-form solutions for optimal resource allocation and task splitting ratios by focusing on the design of uplink communication from UEs to ENs and CP while assuming orthogonal time-division multiple access (TDMA) on wireless access uplink channel and a fixed allocation of fronthaul capacity across the UEs. Reference \cite{Wu-et-al:TVT18} also addressed the design of the task splitting ratios under the assumption that the task of each UE can be split into multiple subtasks that are offloaded to multiple ENs.

In a D-RAN, ENs perform local signal processing for channel encoding and decoding. Thus, the overall performance can be degraded by interference in dense networks. 
In this paper, we propose integrating collaborative fractional cloud-edge offloading within a cloud radio access network (C-RAN) architecture \cite{Simeone-et-al:JCN16}, while accounting for the contributions of both uplink and downlink. 
In a C-RAN, as illustrated in Fig. \ref{fig:system-model}, joint signal processing, in the form of cooperative precoding and detection, at the CP enables effective interference management. Unlike the case of D-RANs, the design of C-RAN systems entails the additional challenge of optimizing the use of ENs-CP \textit{fronthaul} links \cite{Park-et-al:SPM, Zhou-Yu:TSP16, Park-et-al:TWC}.
In this regard, we note that, although fronthaul constraints were also considered in \cite{Ren-et-al:TVT19} for the design within a D-RAN system, a simple data forwarding model was assumed with fixed capacity allocation among the UEs.
In \cite{Garcia-et-al:JSAC18}, the authors tackled the optimization of functional split for collaborative computing systems equipped with a packet-based fronthaul network. However, it was assumed in \cite{Garcia-et-al:JSAC18} that the physical-layer (PHY) functionalities, which include channel encoding and decoding, are located only at ENs.
In \cite{Yang-et-al:IOT}, the authors addressed the task allocation and traffic path planning problem for a C-RAN system under the assumption that the service latency consists of task processing delay and path delay only on fronthaul links.


In this work, we address the optimization of C-RAN signal processing for the purpose of enabling collaborative cloud and edge mobile computing with minimal end-to-end two-way latency. We proceed by first reviewing the design of collaborative cloud and edge computing system within a D-RAN architecture. Unlike \cite{Ren-et-al:TVT19, Zing-et-al:IOT}, which considered one-way uplink design with inter-UE TDMA and fixed fronthaul capacity allocation, we address the design of two-way communications with both TDMA and non-orthogonal multiple access strategies and we treat the fronthaul capacity allocation as optimization variables.
Then, we address the design of C-RAN system for collaborative offloading. For all the design problems, we consider the criterion of minimizing two-way end-to-end latency for computation offloading as in \cite{Ren-et-al:TVT19, Yang-et-al:TC15, Mao-et-al:JSAC16, Yousefpour-et-al:IoT18}. To tackle the formulated problems, which turn out to be non-convex, we adopt fractional programming (FP) and matrix FP \cite{Shen-Yu:TSP18, Shen:TN19}. We present extensive numerical results that confirm the convergence of the proposed optimization algorithms, the advantages of C-RAN architecture as compared to D-RAN \cite{Ren-et-al:TVT19}, and the impact of collaborative cloud and edge computing on latency with C-RAN.

The paper is organized as follows. In Sec. \ref{sec:System-Model}, we describe the system model including the computational tasks, computational capabilities, wireless channel and fronthaul transmission models. In Sec. \ref{sec:D-RAN}, we discuss the design of collaborative cloud and edge mobile computing system within the D-RAN architecture, and the design for a C-RAN system is discussed in Sec. \ref{sec:C-RAN}. We provide extensive numerical results in Sec. \ref{sec:numerical} to validate the performance gain of the proposed architecture as compared to the D-RAN solution. We conclude the paper in Sec. \ref{sec:conclusion}.

\textit{Notations}: We denote the set of all $M\times N$ complex matrices by $\mathbb{C}^{M\times N}$. The notation $\mathbf{x}\sim\mathcal{CN}(\boldsymbol{\mu}, \mathbf{\Omega})$ indicates that $\mathbf{x}$ is a column vector following circularly symmetric complex Gaussian distribution with mean vector $\boldsymbol{\mu}$ and covariance matrix $\mathbf{\Omega}$. We also use the notation $I(\mathbf{x}; \mathbf{y})$ to represent the mutual information between random vectors $\mathbf{x}$ and $\mathbf{y}$. A block diagonal matrix, whose diagonal blocks are given as $\mathbf{A}_1,\ldots,\mathbf{A}_L$, is denoted by 
$\text{diag}(\{\mathbf{A}_l\}_{l\in\{1,\ldots,L\}})$.
Lastly, $\mathbb{E}[\cdot]$ represents the expectation operator, and $||\mathbf{x}||$ denotes the Euclidean 2-norm of a vector $\mathbf{x}$.

\section{System Model\label{sec:System-Model}}

\begin{figure}
\centering\includegraphics[width=9cm,height=7cm,keepaspectratio]{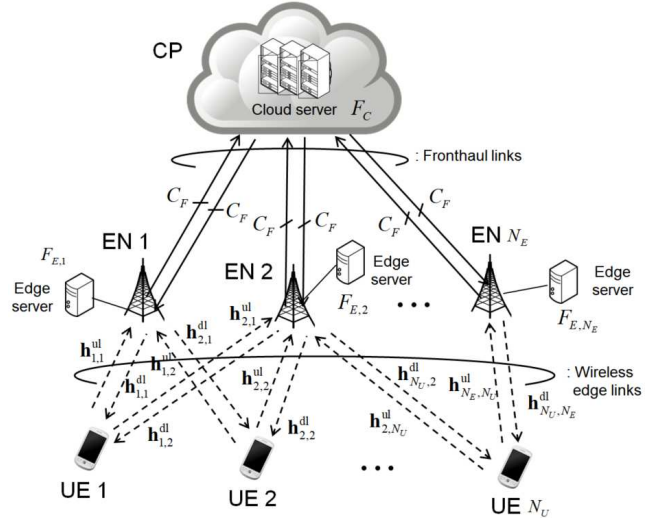}\caption{{\footnotesize{}\label{fig:system-model}Illustration of collaborative cloud and edge mobile computing system within C-RAN architecture.}}
\end{figure}

As illustrated in Fig. \ref{fig:system-model}, we consider a collaborative cloud and edge mobile computing system, in which $N_U$ single-antenna mobile UEs offload their computational tasks to a network consisting of $N_E$ ENs and a CP. 
In order to exchange computational input information, the UEs communicate with the ENs over a wireless uplink channel, and each EN is connected to the CP through dedicated fronthaul link of finite capacity $C_F^{\text{ul}}$ bits per second (bps). For communication in the reverse direction from CP to each EN, the fronthaul has capacity of $C_F^{\text{dl}}$ bps, and the ENs transmit to the UEs in a wireless downlink channel. For convenience, we define the sets $\mathcal{N}_U\triangleq\{1,2,\ldots,N_U\}$ and $\mathcal{N}_E\triangleq\{1,2,\ldots,N_E\}$ of indices of UEs and ENs, respectively.
We denote the number of antennas of EN $i$ as $n_{E,i}$, and the number of all ENs' antennas is $n_{E} = \sum_{i\in\mathcal{N}_E} n_{E,i}$.
The bandwidths of uplink and downlink channels are $W^{\text{ul}}$ and $W^{\text{dl}}$, respectively, which are measured in Hz.

\subsection{Computational Tasks and Collaborative Computing Model} \label{sub:computation-model-task-split}

As in \cite{Ashuwaili:TSIPN17,Ren-et-al:TVT19}, we assume that the UEs have limited computing powers, and hence offload their whole tasks to ENs  or CP without local processing.
We define $b_{I,k}$ and $b_{O,k}$ as the numbers of input and output bits for the task of UE $k$. We assume that $V_k$ CPU cycles are required to process one bit of the task of UE $k$ so that the task of UE $k$ requires $b_{I,k} V_k$ CPU cycles in total.
The computing powers of each EN $i$ and CP are denoted by $F_{E,i}$ and $F_C$, respectively, whose units are CPU cycles per second.

For each UE $k$, we allow for a collaborative cloud and edge computing \cite{Ashuwaili:TSIPN17,Ren-et-al:TVT19}. This means that a part of the task of UE $k$ is processed by a predetermined EN $i_k$, while the rest of the task is offloaded to the CP. 
We define a variable $c_k \in [0,1]$ which controls the fraction of the task of UE $k$ that is processed by EN $i_k$. Accordingly, EN $i_k$ receives the input information of $c_k b_{I,k}$ bits from UE $k$, runs $c_{k} b_{I,k} V_k$ CPU cycles, and reports the resulting output information of $c_k b_{O,k}$ bits back to UE $k$. Similarly, the CP 
receives $(1-c_k) b_{I,k}$ input bits from UE $k$, runs $(1-c_k) b_{I,k} V_k$ CPU cycles, and sends $(1-c_k) b_{O,k}$ output bits to UE $k$.

We define $\mathcal{N}_{U,i}$ as the set of UEs that are associated with EN $i$, i.e., 
\begin{align}
    \mathcal{N}_{U,i} = \big\{ k\in\mathcal{N}_U | i_k = i \big\}. \label{eq:set-UE-closest-to-EN-i}
\end{align}
Therefore, if we denote as $F_{E,i,k}$ the computing power of EN $i$ assigned for UE $k$, the variables $F_{E,i,k}$, $k\in\mathcal{N}_{U,i}$, are subject to the constraint
\begin{align}
    \sum\nolimits_{k\in\mathcal{N}_{U,i}} F_{E,i,k} \leq F_{E,i}. \label{eq:edge-computing-allocation-constraint}
\end{align}
The edge computation latency $\tau_{E,i,k}^{\text{exe}}$ for UE $k$ at EN $i$ with $k\in\mathcal{N}_{U,i}$ is given as
\begin{align}
    \tau_{E,i,k}^{\text{exe}} = \frac{ c_k b_{I,k} V_k }{F_{E,i,k}}. \label{eq:edge-computing-latency-EN-i}
\end{align}

Similarly, denoting the computing power allocated to UE $k$ by the CP as $F_{C,k}$, the variables $F_{C,k}$, $k\in\mathcal{N}_U$, should satisfy the constraint
\begin{align}
    \sum\nolimits_{k\in\mathcal{N}_U} F_{C,k} \leq F_C. \label{eq:cloud-computing-allocation-constraint}
\end{align}
The cloud computing latency $\tau_{C,k}^{\text{exe}}$ for UE $k$ at the CP is given as
\begin{align}
    \tau_{C,k}^{\text{exe}} =  \frac{(1-c_k)b_{I,k}V_k}{F_{C,k}}. \label{eq:cloud-computing-latency-CP}
\end{align}

\subsection{Wireless Channel Model for Edge Link} \label{sub:channel-model}

Assuming the flat fading channel model for both the uplink and downlink wireless edge links, the received signal vector $\mathbf{y}_{i}^{\text{ul}} \in\mathbb{C}^{n_{E,i}\times 1}$ of EN $i$ on the uplink is given as
\begin{align}
    \mathbf{y}_i^{\text{ul}} = \sum\nolimits_{k\in\mathcal{N}_U} \mathbf{h}_{i,k}^{\text{ul}} x^{\text{ul}}_k + \mathbf{z}_i^{\text{ul}}, \label{eq:received-signal-ul}
\end{align}
where $\mathbf{h}^{\text{ul}}_{i,k} \in \mathbb{C}^{n_{E,i}\times 1}$ denotes the channel vector from UE $k$ to EN $i$; $x_k^{\text{ul}}\in\mathbb{C}^{1\times 1}$ indicates the transmit signal of UE $k$; and $\mathbf{z}_i^{\text{ul}}\sim\mathcal{CN}(\mathbf{0}, \sigma_{z,\text{ul}}^2\mathbf{I})$ is the additive noise vector.
Similarly, the received signal $y_k^{\text{dl}} \in \mathbb{C}^{1\times 1}$ of UE $k$ on the downlink can be written as
\begin{align}
    y_{k}^{\text{dl}} = \sum\nolimits_{i\in\mathcal{N}_E} \mathbf{h}_{k,i}^{\text{dl}H} \mathbf{x}_i^{\text{dl}} + z_k^{\text{dl}}, \label{eq:received-signal-dl}
\end{align}
where $\mathbf{h}_{k,i}^{\text{dl}} \in \mathbb{C}^{n_{E,i}\times n_{E,i}}$ represents the channel vector from EN $i$ to UE $k$; $\mathbf{x}^{\text{dl}}_i\in\mathbb{C}^{n_{E,i}\times 1}$ denotes the transmit signal vector of EN $i$; and $z^{\text{dl}}_k\sim\mathcal{CN}(0, \sigma_{z,\text{dl}}^2)$ denotes the additive noise.

The transmit powers of each UE $k$ and EN $i$ are limited as
\begin{align}
    \mathbb{E}\left[|x^{\text{ul}}_k|^2\right] &\leq P^{\text{ul}}, \text{ and } \label{eq:power-constraint-ul} \\
    \mathbb{E}\left[||\mathbf{x}^{\text{dl}}_i||^2\right] &\leq P^{\text{dl}}, \label{eq:power-constraint-dl}
\end{align}
where $P^{\text{ul}}$ and $P^{\text{dl}}$ represent the maximum transmit powers at each UE and EN, respectively.
We define the maximum signal-to-noise ratios (SNRs) of the uplink and downlink channels as $\text{SNR}^{\text{ul}}_{\max} = P^{\text{ul}}/\sigma_{z,\text{ul}}^2$ and $\text{SNR}^{\text{dl}}_{\max} = P^{\text{dl}}/\sigma_{z,\text{dl}}^2$, respectively.
The symbols described in this section are summarized in Table I.

\begin{table}
\centering%
\begin{tabular}{|c|c|}
\hline
Symbol & Meaning\tabularnewline
\hline
\hline
$N_U$, $N_E$ & Numbers of UEs and ENs\tabularnewline
\hline
$\mathcal{N}_{U}$, $\mathcal{N}_{E}$ & Sets of UEs and ENs' indices\tabularnewline
\hline
$n_{E,i}$ & Number of antennas of EN $i$\tabularnewline
\hline
$C_F^{\text{ul}}$, $C_F^{\text{dl}}$ & Capacity of uplink ad downlink fronthaul links\tabularnewline
\hline
$W^{\text{ul}}$, $W^{\text{dl}}$ & Bandwidths of uplink and downlink channels\tabularnewline
\hline
$b_{I,k}$, $b_{O,k}$ & Numbers of input and output bits for UE $k$\tabularnewline
\hline
$V_k$ & Number of CPU cycles per input bit for UE $k$\tabularnewline
\hline
$F_{E,i}$, $F_C$ & CPU frequencies of EN $i$ and CP\tabularnewline
\hline
$c_k$ & Fraction of the task of UE $k$ processed by EN $i_k$\tabularnewline
\hline
$\mathcal{N}_{U,i}$ & Set of UEs associated with EN $i$\tabularnewline
\hline
$P^{\text{ul}}$, $P^{\text{dl}}$ & Maximum transmit powers of each UE and EN\tabularnewline
\hline
$\sigma_{z,\text{ul}}^2$, $\sigma_{z,\text{dl}}^2$ & Noise powers per receive antenna at ENs and UEs\tabularnewline
\hline
$\text{SNR}^{\text{ul}}_{\max}$, $\text{SNR}^{\text{dl}}_{\max}$ & Maximum SNRs of uplink and downlink channels\tabularnewline
\hline
$\mathbf{h}_{i,k}^{\text{ul}}$, $\mathbf{h}_{k,i}^{\text{dl}}$ & Uplink $\&$ downlink channels btw. UE $k$ and EN $i$ \tabularnewline
\hline
$\mathbf{y}_i^{\text{ul}}$, $y_k^{\text{dl}}$ & Received signals of EN $i$ and UE $k$
\tabularnewline
\hline
$x_k^{\text{ul}}$, $\mathbf{x}_i^{\text{dl}}$ & Transmitted signals of UE $k$ and EN $i$
\tabularnewline
\hline
$\mathbf{z}_i^{\text{ul}}$, $z_k^{\text{dl}}$ & Noise signals at EN $i$ and UE $k$
\tabularnewline
\hline
\end{tabular}

~

\centering$\text{{Table\,I:\,\,Table\,\,summarizing\,\,important\,\,symbols\,\,used\,\,throughout\,\,the\,\,paper}}$
\end{table}

\section{Optimization for the D-RAN Architecture} \label{sec:D-RAN}

In this section, we discuss the design of the collaborative cloud and edge mobile computing system under a D-RAN architecture \cite[Sec. III]{Kang-et-al:TWC18}. 
Unlike \cite{Ren-et-al:TVT19}, which considered one-way uplink design with inter-UE TDMA and fixed fronthaul capacity allocation, we address the design of two-way communications with both TDMA and non-orthogonal multiple access strategies while treating the fronthaul capacity allocation as optimization variables.

In D-RAN, each EN $i$ locally decodes the uplink input information transmitted by the associated UEs $\mathcal{N}_{U,i}$ without cooperating with nearby ENs. Also, in the downlink, the computation output information for UEs $\mathcal{N}_{U,i}$ is solely encoded and transmitted by the serving EN $i$.
We discuss the designs with orthogonal TDMA and non-orthogonal multiple access strategies in Sec. \ref{sub:orthogonal-TDMA-access} and \ref{sub:non-orthogonal-access}, respectively.

\subsection{Orthogonal TDMA} \label{sub:orthogonal-TDMA-access}

With TDMA, $N_U$ UEs communicate with $N_E$ ENs on the wireless edge link while being assigned different time slots so that there is no inter-UE interference on wireless channel.
We define $u_k^{\text{ul}} \in [0,1]$ and $u_k^{\text{dl}} \in [0,1]$ as the uplink and downlink time fractions allocated to UE $k$. Thus, the defined fraction variables $\mathbf{u} \triangleq \{u_k^{\text{ul}}, u_k^{\text{dl}}\}_{k\in\mathcal{N}_U}$ should satisfy the constraint
\begin{align}
    \sum\nolimits_{k\in\mathcal{N}_U} u_k^{\text{ul}} = \sum\nolimits_{k\in\mathcal{N}_U} u_k^{\text{dl}} = 1. \label{eq:time-fraction-constraint}
\end{align}

In the uplink, UE $k$ transmits a baseband signal which encodes the input information for its task. Assuming that Gaussian channel codebooks are used, the transmitted signal $x_k^{\text{ul}}$ of UE $k$ is distributed as $x_k^{\text{ul}} \sim \mathcal{CN}(0,p_k^{\text{ul}})$. Since there is no co-channel interference with orthogonal TDMA, the transmit power $p_k$ of UE $k$ is set to $p_k^{\text{ul}} = P^{\text{ul}}$ without loss of optimality.


With the described transmission model, the achievable data rate $R_k^{\text{ul}}$ between UE $k$ and EN $i$ in the uplink channel is given as $R_k^{\text{ul}} = u_k^{\text{ul}} W^{\text{ul}} I(x_k^{\text{ul}} ; \mathbf{y}_i^{\text{ul}})$, where the mutual information $I(x_k^{\text{ul}} ; \mathbf{y}_i^{\text{ul}})$ is calculated as
\begin{align}
    I\left(x_k^{\text{ul}} ; \mathbf{y}_i^{\text{ul}}\right) = \log_2 \left( 1 + \left(P^{\text{ul}}/\sigma_{z,\text{ul}}^2\right) \left\Vert\mathbf{h}_{i,k}^{\text{ul}}\right\Vert^2\right). \label{eq:rate-uplink-TDMA}
\end{align}
The uplink latency $\tau_{E,k}^{\text{ul}}$ on the wireless edge link for UE $k$ is then given as
\begin{align}
    \tau_{E,k}^{\text{ul}} = \frac{ b_{I,k} }{R_k^{\text{ul}}}. \label{eq:edge-latency-uplink-TDMA}
\end{align}

Among the received $b_{I,k}$ bits from UE $k\in\mathcal{N}_{U,i}$, EN $i$ processes only $c_k b_{I,k}$ bits using its edge server and forwards the remaining $(1-c_k) b_{I,k}$ bits to the CP on the fronthaul link for cloud computing. We denote the partial capacity of the fronthaul link between EN $i$ and CP that is used for transferring the $(1-c_k) b_{I,k}$ input bits for UE $k$ by $C_{F,k}^{\text{ul}} \geq 0$ so that $C_{F,k}^{\text{ul}}$, $k\in\mathcal{N}_{U,i}$, satisfy the constraint
\begin{align}
    \sum\nolimits_{k\in\mathcal{N}_{U,i}} C_{F,k}^{\text{ul}} \leq C_F^{\text{ul}},
\end{align}
for all $i\in\mathcal{N}_E$. 
For given $C_{F,k}^{\text{ul}}$, the uplink fronthaul latency $\tau_{F,k}^{\text{ul}}$ of UE $k$ is given as
\begin{align}
    \tau_{F,k}^{\text{ul}} = \frac{(1-c_k)b_{I,k}}{C_{F,k}^{\text{ul}}}. \label{eq:latency-fronthaul-uplink-TDMA}
\end{align}

The CP processes the received $(1-c_k)b_{I,k}$ bits for UE $k$ producing output information of $(1-c_k)b_{O,k}$ bits.
The output bits are transmitted to EN $i_k$ that serves UE $k$.
We denote by $C_{F,k}^{\text{dl}} \geq 0$ the partial capacity of the fronthaul link from CP to EN $i_k$ that is used to transfer the $(1-c_k)b_{O,k}$ bits for UE $k$. Thus, the following constraint should be satisfied:
\begin{align}
    \sum\nolimits_{ k\in\mathcal{N}_{U,i} } C_{F,k}^{\text{dl}} \leq C_F^{\text{dl}},
\end{align}
for all $i\in\mathcal{N}_E$.
The downlink fronthaul latency $\tau_{F,k}^{\text{dl}}$ of UE $k$ for given $C_F^{\text{dl}}$ is given as
\begin{align}
    \tau_{F,k}^{\text{dl}} = \frac{ (1-c_k) b_{O,k} }{C_{F,k}^{\text{dl}}}. \label{eq:latency-fronthaul-downlink-TDMA}
\end{align}

In the downlink, each EN $i$ reports the computation output information of $b_{O,k}$ bits to UE $k\in \mathcal{N}_{U,i}$. To this end, EN $i$ encodes the output information with Gaussian channel codebook producing an encoded baseband signal $\mathbf{s}_{k}^{\text{dl}} \sim \mathcal{CN}(\mathbf{0}, \mathbf{Q}_k^{\text{dl}})$ with $\mathbb{E}[||\mathbf{x}_k^{\text{dl}}||^2] = \text{tr}(\mathbf{Q}_k^{\text{dl}}) \leq P^{\text{dl}}$. Therefore, EN $i$ transmits the encoded signal $\mathbf{s}_{k}^{\text{dl}}$ during a fraction $u_k^{\text{dl}}$ of the downlink time slot.
For given $\mathbf{Q}_k^{\text{dl}}$, the achievable downlink data rate $R_k^{\text{dl}}$ is given as $R_k^{\text{dl}} = u_k^{\text{dl}} W^{\text{dl}} I( \mathbf{s}_k^{\text{dl}}; y_k^{\text{dl}} )$ with $I( \mathbf{s}_k^{\text{dl}}; y_k^{\text{dl}} )$ computed as
\begin{align}
    I\left( \mathbf{s}_k^{\text{dl}}; y_k^{\text{dl}} \right) = \log_2\left( 1 + \left(1/\sigma_{z,\text{dl}}^2\right) \mathbf{h}_{k,i}^{\text{dl} H} \mathbf{Q}_k^{\text{dl}} \mathbf{h}_{k,i}^{\text{dl}} \right). \label{eq:mutual-information-downlink-TDMA}
\end{align}
The optimal covariance matrix $\mathbf{Q}_k^{\text{dl}\star}$, that maximizes the mutual information in (\ref{eq:mutual-information-downlink-TDMA}) while satisfying the constraint $\text{tr}(\mathbf{Q}_k^{\text{dl}}) \leq P^{\text{dl}}$, implements conjugate beamforming \cite{Lo:TCOM} and is given as
\begin{align}
    \mathbf{Q}_k^{\text{dl}\star} = P^{\text{dl}} \tilde{\mathbf{h}}_{k,i}^{\text{dl}} \tilde{\mathbf{h}}_{k,i}^{\text{dl}H}, \label{eq:optimal-covariance-downlink-TDMA}
\end{align}
where $\tilde{\mathbf{h}}_{k,i}^{\text{dl}} = \mathbf{h}_{k,i}^{\text{dl}} / ||\mathbf{h}_{k,i}^{\text{dl}}||$. By substituting (\ref{eq:optimal-covariance-downlink-TDMA}) into (\ref{eq:mutual-information-downlink-TDMA}), we obtain the maximized mutual information value $I( \mathbf{s}_k^{\text{dl}}; y_k^{\text{dl}} )$ as
\begin{align}
    I\left( \mathbf{s}_k^{\text{dl}}; y_k^{\text{dl}} \right) = \log_2\left( 1 + \left(P^{\text{dl}} / \sigma_{z,\text{dl}}^2\right) \left\Vert \mathbf{h}_{k,i}^{\text{dl}} \right\Vert^2 \right).
\end{align}
The downlink latency $\tau_{E,k}^{\text{dl}}$ for UE $k$ on the wireless edge link is hence given as
\begin{align}
    \tau_{E,k}^{\text{dl}} = \frac{ b_{O,k} }{R_{k}^{\text{dl}}}. \label{eq:edge-latency-downlink-TDMA}
\end{align}

Finally, the overall latency $\tau_{T,k}$ for each UE $k$ is given as
\begin{align}
    \tau_{T,k} = \tau_{E,k}^{\text{ul}} + \max\big\{\tau_{E,i_k,k}^{\text{exe}}, \tau_{F,k}^{\text{ul}} + \tau_{C,k}^{\text{exe}} + \tau_{F,k}^{\text{dl}}\big\} + \tau_{E,k}^{\text{dl}},
\end{align}
where the second term indicates that local edge computing at EN $i_k$ and fronthaul transmissions can take place simultaneously.
As a result, the total latency required for completing the tasks of all the participating UEs is given as
\begin{align}
    \tau_T = \max_{k\in\mathcal{N}_U} \tau_{T,k}. \label{eq:total-latency-DRAN}
\end{align}

We tackle the problem of optimizing the variables $\mathbf{c} \triangleq \{c_k\}_{k\in\mathcal{N}_U}$, $\mathbf{u}$, $\mathbf{F} \triangleq \{F_{E,i,k}\}_{i\in\mathcal{N}_E, k\in\mathcal{N}_{U,i}} \cup \{F_{C,k}\}_{k\in\mathcal{N}_U}$ and $\mathbf{C}_F \triangleq \{ C_{F,k}^{\text{ul}}, C_{F,k}^{\text{dl}} \}_{k\in\mathcal{N}_U}$ with the goal of minimizing the total latency $\tau_T$. We formulate this problem as
\begin{subequations} \label{eq:problem-TDMA}
\begin{align}
    \!\!\!\!\!\!\!\!\!\!\!\!\!\!\!\!\!\!\!\!\!\!\!\!\!\underset{ ^{\mathbf{c}\geq 0, \mathbf{u} \geq 0, \mathbf{F} \geq 0,} _{ \,\,\,\,\,\,\,\,\mathbf{C}_F \geq 0, \boldsymbol{\tau}}  }{\mathrm{minimize}}\,\, &  \max_{k\in\mathcal{N_U}} \tau_{T,k} \label{eq:problem-TDMA-cost} \\
    \mathrm{s.t.} \,\,\,\,\,\,\, & \tau_{E,k}^{\text{ul}} \geq \frac{b_{I,k}}{ u_k^{\text{ul}} \tilde{R}_k^{\text{ul}} }, \,\, k\in\mathcal{N}_U, \label{eq:problem-TDMA-latency-edge-uplink} \\
    & \tau_{F,k}^{\text{ul}} \geq \frac{ (1-c_k)b_{I,k} }{ C_{F,k}^{\text{ul}} }, \,\, k\in\mathcal{N}_U, \label{eq:problem-TDMA-latency-fronthaul-uplink} \\
    & \tau_{E,k}^{\text{dl}} \geq \frac{b_{O,k}}{u_k^{\text{dl}} \tilde{R}_k^{\text{dl}}}, \,\, k\in\mathcal{N}_U, \label{eq:problem-TDMA-latency-edge-downlink} \\
    & \tau_{F,k}^{\text{dl}} \geq \frac{ (1-c_k) b_{O,k} }{C_{F,k}^{\text{dl}}} \,\, k\in\mathcal{N}_U, \label{eq:problem-TDMA-latency-fronthaul-downlink} \\
    & \tau_{E,i_k,k}^{\text{exe}} \geq \frac{c_k b_{I,k} V_k}{ F_{E,i_k,k} }, \,\, k\in\mathcal{N}_U, \label{eq:problem-TDMA-latency-exe-edge} \\
    & \tau_{C,k}^{\text{exe}} \geq \frac{(1-c_k)b_{I,k}V_k}{F_{C,k}}, \,\, k\in\mathcal{N}_U, \label{eq:problem-TDMA-latency-exe-cloud}\\
    & c_k \in [0,1], \,\, k\in\mathcal{N}_U, \\
    & \sum\nolimits_{k\in\mathcal{N}_U} u_k^{\text{ul}} = \sum\nolimits_{k\in\mathcal{N}_U} u_k^{\text{dl}} = 1, \\
    & \sum\nolimits_{k\in\mathcal{N}_{U,i}} F_{E,i,k} \leq F_{E,i}, \,\, i\in\mathcal{N}_E, \label{eq:problem-TDMA-computing-allocation-edge}\\
    & \sum\nolimits_{k\in\mathcal{N}_U} F_{C,k} \leq F_C, \label{eq:problem-TDMA-computing-allocation-cloud} \\
    & \sum\nolimits_{k\in\mathcal{N}_{U,i}} C_{F,k}^{\text{ul}} \leq C_F^{\text{ul}}, \,\, i\in\mathcal{N}_E, \label{eq:problem-sum-CF-uplink} \\
    & \sum\nolimits_{k\in\mathcal{N}_{U,i}} C_{F,k}^{\text{dl}} \leq C_F^{\text{dl}}, \,\, i\in\mathcal{N}_E, \label{eq:problem-sum-CF-downlink}
\end{align}
\end{subequations}
with the notations $\tilde{R}_k^{\text{ul}} = W^{\text{ul}} I(x_k^{\text{ul}} ; \mathbf{y}_{i_k}^{\text{ul}})$,  $\tilde{R}_k^{\text{dl}} = W^{\text{dl}} I(\mathbf{s}_k^{\text{dl}} ; y_k^{\text{dl}})$, and $\boldsymbol{\tau} = \{ \tau_{E,k}^{\text{ul}}, \tau_{F,k}^{\text{ul}} , \tau_{E,k}^{\text{dl}}, \tau_{F,k}^{\text{dl}} , \tau_{E,i_k,k}^{\text{exe}}, \tau_{C,k}^{\text{exe}} \}_{k\in\mathcal{N}_U}$.

The problem (\ref{eq:problem-TDMA}) is non-convex due to the constraints (\ref{eq:problem-TDMA-latency-fronthaul-uplink}) and (\ref{eq:problem-TDMA-latency-fronthaul-downlink})-(\ref{eq:problem-TDMA-latency-exe-cloud}).  
We can tackle the non-convex problem by coordinate descent approach \cite[Sec. 1.8]{Bertsekas}, since the problem becomes convex if we fix one of the variable sets $\mathbf{c}$ and $\{\mathbf{F}, \mathbf{C}_F\}$.
However, the coordinate descent approach cannot be directly applied to the problems that will be discussed in Sec. \ref{sub:non-orthogonal-access} and \ref{sec:C-RAN}, and hence we consider FP \cite{Shen-Yu:TSP18} as a solution method, which can overcome this limitation.

We observe that all the constraints (\ref{eq:problem-TDMA-latency-fronthaul-uplink}) and (\ref{eq:problem-TDMA-latency-fronthaul-downlink})-(\ref{eq:problem-TDMA-latency-exe-cloud}), that induce the non-convexity of the problem (\ref{eq:problem-TDMA}), can be expressed as a function of ratios of optimization variables. 
It was shown in \cite{Shen-Yu:TSP18} that FP is suitable for approximating those constraints by convex constraints.
In more detail,
based on \cite[Cor. 1]{Shen-Yu:TSP18}, we can show that, for any real values $\lambda_{F,k}^{\text{ul}}$, $\lambda_{F,k}^{\text{dl}}$, $\lambda_{E,i_k,k}^{\text{exe}}$ and $\lambda_{C,k}^{\text{exe}}$, the following constraints are stricter than (\ref{eq:problem-TDMA-latency-fronthaul-uplink}) and (\ref{eq:problem-TDMA-latency-fronthaul-downlink})-(\ref{eq:problem-TDMA-latency-exe-cloud}):
\begin{subequations} \label{eq:rewrite-latency-constraint-TDMA}
\begin{align}
    2\lambda_{F,k}^{\text{ul}} \sqrt{\tau_{F,k}^{\text{ul}}} - (\lambda_{F,k}^{\text{ul}})^2 \left(1-c_k\right) & \geq \frac{b_{I,k}}{C_{F,k}^{\text{ul}}}, \,\,  k\in\mathcal{N}_U, \label{eq:rewrite-latenccy-constraint-TDMA-1} \\
    2\lambda_{F,k}^{\text{dl}} \sqrt{\tau_{F,k}^{\text{dl}}} - (\lambda_{F,k}^{\text{dl}})^2 \left(1-c_k\right) & \geq \frac{b_{O,k}}{C_{F,k}^{\text{dl}}}, \,\, k\in \mathcal{N}_U, \\
    2\lambda_{E,i_k,k}^{\text{exe}} \sqrt{ \tau_{E,i_k,k}^{\text{exe}} } - (\lambda_{E,i_k,k}^{\text{exe}})^2 c_k  & \geq \frac{b_{I,k} V_k}{F_{E,i_k,k}},  k\in\mathcal{N}_U, \label{eq:rewrite-latency-constraint-TDMA-3} \\
    2 \lambda_{C,k}^{\text{exe}} \sqrt{\tau_{C,k}^{\text{exe}}} - (\lambda_{C,k}^{\text{exe}})^2 \left(1-c_k\right) & \geq \frac{b_{I,k}V_k}{ F_{C,k} }, \,\, k\in\mathcal{N}_U. \label{eq:rewrite-latenccy-constraint-TDMA-4}
\end{align}
\end{subequations}
The above constraints have the following desirable properties: they are convex constraints, if the auxiliary variables $\lambda_{F,k}^{\text{ul}}$, $\lambda_{F,k}^{\text{dl}}$, $\lambda_{E,i_k,k}^{\text{exe}}$ and $\lambda_{C,k}^{\text{exe}}$ are fixed. And they become equivalent to  (\ref{eq:problem-TDMA-latency-fronthaul-uplink}) and (\ref{eq:problem-TDMA-latency-fronthaul-downlink})-(\ref{eq:problem-TDMA-latency-exe-cloud}), if the variables $\lambda_{F,k}^{\text{ul}}$, $\lambda_{F,k}^{\text{dl}}$, $\lambda_{E,i_k,k}^{\text{exe}}$ and $\lambda_{C,k}^{\text{exe}}$ are given as
\begin{align}
    & \lambda_{F,k}^{\text{ul}} = \frac{ \sqrt{\tau_{F,k}^{\text{ul}}} }{1-c_k}, \,
    \lambda_{F,k}^{\text{dl}} = \frac{\sqrt{\tau_{F,k}^{\text{dl}}}}{1 - c_k}, \,
    \lambda_{E,i_k,k}^{\text{exe}} = \frac{ \sqrt{\tau_{E,i_k,k}^{\text{exe}}} }{ c_k }, \nonumber \\
    &\text{and }
    \lambda_{C,k}^{\text{exe}} = \frac{ \sqrt{\tau_{C,k}^{\text{exe}}} }{1-c_k}. \label{eq:update-lambda-TDMA}
\end{align}

Based on the above observation, we consider the problem obtained by replacing the constraints (\ref{eq:problem-TDMA-latency-fronthaul-uplink}) and (\ref{eq:problem-TDMA-latency-fronthaul-downlink})-(\ref{eq:problem-TDMA-latency-exe-cloud}) with (\ref{eq:rewrite-latency-constraint-TDMA}) in (\ref{eq:problem-TDMA}) and adding $\boldsymbol{\lambda}=\{ \lambda_{F,k}^{\text{ul}}, \lambda_{F,k}^{\text{dl}}, \lambda_{E,i_k,k}^{\text{exe}}, \lambda_{C,k}^{\text{exe}} \}_{k\in\mathcal{N}_U}$ as optimization variables.
To tackle the obtained problem, which has the same optimal value as (\ref{eq:problem-TDMA}), we propose an iterative algorithm, in which the variables $\{\mathbf{c}, \mathbf{u}, \mathbf{F}, \mathbf{C}_F, \boldsymbol{\tau}\}$ and $\boldsymbol{\lambda}$ are alternately updated. Since the optimization of $\{\mathbf{c}, \mathbf{u}, \mathbf{F}, \mathbf{C}_F, \boldsymbol{\tau}\}$ for fixed $\boldsymbol{\lambda}$ is a convex problem, standard convex solvers, such as the CVX software \cite{CVX}, can be used. The optimal $\boldsymbol{\lambda}$ for fixed $\{\mathbf{c}, \mathbf{u}, \mathbf{F}, \mathbf{C}_F, \boldsymbol{\tau}\}$ can be obtained as (\ref{eq:update-lambda-TDMA}), which make the constraints (\ref{eq:rewrite-latenccy-constraint-TDMA-1})-(\ref{eq:rewrite-latenccy-constraint-TDMA-4}) equivalent to the original constraints (\ref{eq:problem-TDMA-latency-fronthaul-uplink}) and (\ref{eq:problem-TDMA-latency-fronthaul-downlink})-(\ref{eq:problem-TDMA-latency-exe-cloud}).
We describe the detailed algorithm in Algorithm 1.

The convex problem solved at Step 4 of each $t$th iteration in Algorithm 1 has stricter constraints than the original problem (\ref{eq:problem-TDMA}). Also, the feasible space of the convex problem contains the solution obtained at the $(t-1)$th iteration. Thus, the solution of the convex problem at the $t$th iteration belongs to the feasible space of problem (\ref{eq:problem-TDMA}) and achieves a lower latency value than the solution of the $(t-1)$th iteration. Therefore, Algorithm 1 produces monotonically decreasing latency values with respect to the iteration index $t$ so that it converges to a locally optimal point. For more formal proof of the convergence of SCA and FP algorithms, we refer to \cite{Scutari-et-al:TSP17:Part1, Shen-Yu:TSP18}.
We can operate Algorithm 1 with an arbitrary initial point that satisfies the conditions (\ref{eq:problem-TDMA-latency-edge-uplink})-(\ref{eq:problem-sum-CF-downlink}). In the simulation section, we initialize the variables $\{\mathbf{c}, \mathbf{u}, \mathbf{F}, \mathbf{C}_F\}$ at Step 1 as 
\begin{subequations}
\begin{align}
& u_k\leftarrow 1/N_U, \, k\in\mathcal{N}_U, \label{eq:init-uk} \\
& c_k\leftarrow 1/2, \, k\in\mathcal{N}_U, \label{eq:init-ck} \\
& F_{E,i,k} \leftarrow F_{E,i} / |\mathcal{N}_{U,i}|, \, k\in\mathcal{N}_{U,i}, i\in\mathcal{N}_E, \label{eq:init-FE-k} \\
& F_{C,k} \leftarrow F_{C} / N_U, \, k\in\mathcal{N}_{U}, \label{eq:init-FC-k} \\
& C_{F,k}^{m} \leftarrow C_F^{\text{ul}} / |\mathcal{N}_{U,i}|, \, k\in\mathcal{N}_{U,i}, i\in\mathcal{N}_E, m\in\{\text{ul},\text{dl}\}. \label{eq:init-CF}
\end{align}
\end{subequations}
For the given $\{\mathbf{c}, \mathbf{u}, \mathbf{F}, \mathbf{C}_F\}$, we compute an initial value for  $\boldsymbol{\tau}$ according to (\ref{eq:edge-latency-uplink-TDMA}), (\ref{eq:latency-fronthaul-uplink-TDMA}), (\ref{eq:latency-fronthaul-downlink-TDMA}), and (\ref{eq:edge-latency-downlink-TDMA}).

\begin{algorithm}
\caption{Alternating optimization algorithm that tackles problem  (\ref{eq:problem-TDMA})}

\textbf{1.} Initialize $\{\mathbf{c}, \mathbf{u}, \mathbf{F}, \mathbf{C}_F, \boldsymbol{\tau}\}$ as arbitrary values that satisfy the constraints (\ref{eq:problem-TDMA-latency-edge-uplink})-(\ref{eq:problem-sum-CF-downlink}), and set $t\leftarrow 1$.

\textbf{2.} Calculate the total latency $\tau_T$ in (\ref{eq:total-latency-DRAN}) with the initialized $\{\mathbf{c}, \mathbf{u}, \mathbf{F}, \mathbf{C}_F, \boldsymbol{\tau}\}$, and set $\tau_T^{(0)}\leftarrow \tau_T$.

\textbf{3.} Set $\boldsymbol{\lambda}$ according to (\ref{eq:update-lambda-TDMA}).

\textbf{4.} Update the variables $\{\mathbf{c}, \mathbf{u}, \mathbf{F}, \mathbf{C}_F, \boldsymbol{\tau}\}$ as a solution of the convex problem which is obtained by replacing the constraints (\ref{eq:problem-TDMA-latency-fronthaul-uplink}) and (\ref{eq:problem-TDMA-latency-fronthaul-downlink})-(\ref{eq:problem-TDMA-latency-exe-cloud}) with (\ref{eq:rewrite-latenccy-constraint-TDMA-1})-(\ref{eq:rewrite-latenccy-constraint-TDMA-4}) and then by fixing $\boldsymbol{\lambda}$.

\textbf{5.} Calculate the total latency $\tau_T$ with the updated $\{\mathbf{c}, \mathbf{u}, \mathbf{F}, \mathbf{C}_F, \boldsymbol{\tau}\}$, and set  $\tau_T^{(t)}\leftarrow \tau_T$.

\textbf{6.} Stop if $|\tau_T^{(t)} - \tau_T^{(t-1)}|\leq \delta$ or $t>t_{\max}$. Otherwise, set $t\leftarrow t+1$ and go back to Step 2.

\end{algorithm}

The complexity of Algorithm 1 is given by the number of iterations multiplied by the complexity of solving the convex problem at each iteration (i.e., Step 4). 
The complexity of solving a generic convex problem is upper bounded by $\mathcal{O}( n(n^3+M) \log(1/\epsilon) )$ \cite[p. 4]{BTal-Nemirovski}, where $n$ denotes the number of optimization variables, $M$ is the number of arithmetic operations required to compute the objective and constraint functions, and $\epsilon$ represents the desired error tolerance. The numbers $n$ and $M$ equal $n=13 N_U$ and $M = 45 N_U$, respectively, for the convex problem solved at Step 4 of Algorithm 1. 
However, to the best of our knowledge, the analysis of the convergence rate of general SCA algorithms is still an open problem. Instead, we provide some numerical evidence of the fast convergence of Algorithm 1 in Sec. \ref{sec:numerical}.


\subsection{Non-Orthogonal Multiple Access} \label{sub:non-orthogonal-access}

In this subsection, we discuss the design with non-orthogonal multiple access. With non-orthogonal access, $N_U$ UEs communicate simultaneously with $N_E$ ENs on the same time and frequency resource. Therefore, the uplink and downlink communications on the wireless edge link are impaired by inter-UE interference signals, while benefiting from transmission on a larger time interval. The computation and fronthaul transmission models are the same as the one described in Sec. \ref{sub:orthogonal-TDMA-access}, and we detail here only the uplink and downlink communication phases and the resulting latency performance.

As in Sec. \ref{sub:orthogonal-TDMA-access}, we assume that each UE $k$ uses a Gaussian channel codebook so that its transmitted signal $x_k^{\text{ul}}$ is distributed as $x_k^{\text{ul}} \sim \mathcal{CN}(0,p_k^{\text{ul}})$. The transmit power $p_k^{\text{ul}}$ is subject to the constraint $p_k^{\text{ul}} \in [0, P^{\text{ul}}]$. 
Due to the presence of inter-UE interference signals,
full power transmission at all UEs may cause an optimality loss. This suggests that we need to carefully design the transmit power variables $p_k^{\text{ul}}$, $k\in\mathcal{N}_U$, by adapting to channel state information (CSI).

Each EN $i$ needs to decode the signals $\{x_k^{\text{ul}}\}_{k\in\mathcal{N}_{U,i}}$ based on the received signal $\mathbf{y}_i^{\text{ul}}$.
We assume that the signals $\{x_k^{\text{ul}}\}_{k\in\mathcal{N}_{U,i}}$ are detected in parallel without successive interference cancellation (SIC) as in \cite{Inaltekin-Hanly:TIT12, Joudeh-Clerckx:TIT19} in order to minimize the decoding delay. We leave the design and analysis with SIC decoding \cite{Yang-et-al:TCOM12} while taking into account the decoding delay for future work.

Under the assumption of parallel decoding, the achievable rate $R_k^{\text{ul}}$ of UE $k$ in the uplink channel is given as $R_k^{\text{ul}} = W^{\text{ul}} I(x_k^{\text{ul}}; \mathbf{y}_{i_k}^{\text{ul}})$ with the mutual information value computed as
\begin{align}
    & I(x_k^{\text{ul}}; \mathbf{y}_{i_k}^{\text{ul}}) = f_{E,k}^{\text{ul}}\left(\mathbf{p}\right) = \label{eq:rate-edge-uplink-NOMA} \\
    & \Psi\left(  p_k^{\text{ul}} \mathbf{h}_{i_k,k}^{\text{ul}} \mathbf{h}_{i_k,k}^{\text{ul}\,H} \, \boldsymbol{,} \,\, \sigma_{z,\text{ul}}^2 \mathbf{I} + \sum\nolimits_{l\in\mathcal{N}_U \setminus \{k\}} \!\!\! p_l^{\text{ul}} \mathbf{h}_{i_k,l}^{\text{ul}} \mathbf{h}_{i_k,l}^{\text{ul}\,H} \right). \nonumber 
\end{align}
Here we have defined the notation $\mathbf{p}\triangleq \{ p_k^{\text{ul}} \}_{k\in\mathcal{N}_U}$, and the function
\begin{align}
    \Psi\left(\mathbf{A}, \mathbf{B}\right) = \log_2\det\left(\mathbf{I} + \mathbf{B}^{-1}\mathbf{A}\right)
\end{align}
For given $R_k^{\text{ul}}$, the uplink edge latency $\tau_{E,k}^{\text{ul}}$ for UE $k$ is given as (\ref{eq:edge-latency-uplink-TDMA}).

For the downlink edge link, each EN $i$ transmits a superposition of the signals $\mathbf{s}_k^{\text{dl}}$, $k\in\mathcal{N}_{U,i}$, where $\mathbf{s}_k^{\text{dl}}\sim\mathcal{CN}(\mathbf{0}, \mathbf{Q}_k^{\text{dl}})$ encodes the task output of UE $k$. The transmit signal of EN $i$ is written as
\begin{align}
    \mathbf{x}_i^{\text{dl}} = \sum\nolimits_{k\in\mathcal{N}_{U,i}} \mathbf{s}_k^{\text{dl}}. \label{eq:superposition-TDMA-downlink}
\end{align}
With the above transmission model, the downlink transmit power constraint (\ref{eq:power-constraint-dl}) can be expressed as $\sum_{k\in\mathcal{N}_{U,i}} \text{tr}( \mathbf{Q}_k^{\text{dl}} ) \leq P^{\text{dl}}$, and the achievable rate $R_k^{\text{dl}}$ of UE $k$ on the wireless edge link is given as $R_k^{\text{dl}} = W^{\text{dl}} I(\mathbf{s}_k^{\text{dl}} ; y_k^{\text{dl}})$ with
\begin{align}
    & I\left( \mathbf{s}_k^{\text{dl}} ; y_k^{\text{dl}} \right) = f_{E,k}^{\text{dl}}\left( \mathbf{Q} \right) = \label{eq:rate-edge-downlink-NOMA}
    \\ 
    & \Psi\left( \mathbf{h}_{k,i_k}^{\text{dl}\,H}\mathbf{Q}_k^{\text{dl}}\mathbf{h}_{k,i_k}^{\text{dl}} \,\boldsymbol{,} \,\, \sigma_{z,\text{dl}}^2 + \sum\nolimits_{l\in\mathcal{N}_U\setminus \{k\}} \mathbf{h}_{k,i_l}^{\text{dl}\,H}\mathbf{Q}_l^{\text{dl}}\mathbf{h}_{k,i_l}^{\text{dl}} \right), \nonumber
\end{align}
where $\mathbf{Q} \triangleq \{ \mathbf{Q}_k^{\text{dl}} \}_{k\in\mathcal{N}_U}$.
For given $R_k^{\text{dl}}$, the downlink edge latency $\tau_{E,k}^{\text{dl}}$ of UE $k$ is given as (\ref{eq:edge-latency-downlink-TDMA}).

For the non-orthogonal multiple access scheme as described above, we aim at jointly optimizing the variables $\mathbf{p}$, $\mathbf{Q}$, $\mathbf{c}$, $\mathbf{F}$ and $\mathbf{C}_F$ with the goal of minimizing the total latency $\tau_T$ in (\ref{eq:total-latency-DRAN}). The problem can be written as
\begin{subequations} \label{eq:problem-NOMA}
\begin{align}
    \!\!\!\!\!\!\!\!\!\!\!\!\!\!\!\!\!\!\!\!\!\!\!\!\!\!\!\!\!\!\!\!\!\!\!\!\!\!\!\!\!\!\!\underset{ ^{\mathbf{p}\geq 0, \mathbf{Q}\succeq\mathbf{0}, \mathbf{c}\geq 0,} _{\mathbf{F} \geq 0, \mathbf{C}_F \geq 0, \boldsymbol{\tau}, \mathbf{R}} }{\mathrm{minimize}}\,\, &  \max_{k\in\mathcal{N_U}} \tau_{T,k} \label{eq:problem-NOMA-cost} \\
    \mathrm{s.t.} \,\,\,\,\,\,\, & \tau_{E,k}^{\text{ul}} \geq \frac{b_{I,k}}{ R_k^{\text{ul}} }, \,\, k\in\mathcal{N}_U, \label{eq:problem-NOMA-latency-edge-uplink} \\
    & \tau_{E,k}^{\text{dl}} \geq \frac{b_{O,k}}{ R_k^{\text{dl}}}, \,\, k\in\mathcal{N}_U, \label{eq:problem-NOMA-latency-edge-downlink} \\
    & \text{(\ref{eq:problem-TDMA-latency-fronthaul-uplink}), (\ref{eq:problem-TDMA-latency-fronthaul-downlink})-(\ref{eq:problem-TDMA-latency-exe-cloud})}, \label{eq:problem-NOMA-latency-non-convex} \\
    & R_{E,k}^{\text{ul}} \leq f_{E,k}^{\text{ul}}\left(\mathbf{p}\right), \,\, k\in\mathcal{N}_U, \label{eq:problem-NOMA-rate-edge-uplink} \\
    & R_{E,k}^{\text{dl}} \leq f_{E,k}^{\text{dl}} \left(\mathbf{Q}\right), \,\, k\in\mathcal{N}_U, \label{eq:problem-NOMA-rate-edge-downlink} \\
    & p_k^{\text{ul}} \leq P^{\text{ul}}, \,\, k\in\mathcal{N}_U, \label{eq:problem-NOMA-power-uplink} \\
    & \sum\nolimits_{k\in\mathcal{N}_{U,i}} \text{tr}\left(\mathbf{Q}_k^{\text{dl}}\right) \leq P^{\text{dl}}, \,\, i\in\mathcal{N}_E, \label{eq:problem-NOMA-power-downlink} \\
    & c_k \in [0,1], \,\, k\in\mathcal{N}_U, \\
    & \text{(\ref{eq:problem-TDMA-computing-allocation-edge})-(\ref{eq:problem-sum-CF-downlink})}, \label{eq:problem-NOMA-sum-F-C}
\end{align}
\end{subequations}
where we have defined $\mathbf{R} \triangleq \{ R_{E,k}^{\text{ul}}, R_{E,k}^{\text{dl}} \}_{k\in\mathcal{N}_U}$.


We note that it is more challenging to tackle problem (\ref{eq:problem-NOMA}) than (\ref{eq:problem-TDMA}) due to the presence of inter-UE interference signals on the wireless edge links.
Accordingly, the uplink and downlink transmission strategies on edge links, which are characterized by the variables $\mathbf{p}$ and $\mathbf{Q}$, need to be jointly optimized. Also, the constraints (\ref{eq:problem-NOMA-rate-edge-uplink}) and (\ref{eq:problem-NOMA-rate-edge-downlink}) on the edge throughputs, which involve matrix variables $\mathbf{Q}$, are not convex.
To address these complications, we employ FP \cite{Shen-Yu:TSP18} as well as matrix FP \cite{Shen:TN19}, which is a generalized version of \cite{Shen-Yu:TSP18}.


We first observe that the constraints (\ref{eq:problem-NOMA-latency-non-convex}), that are expressed as a function of ratios of scalar optimization variables, can be handled by  FP \cite{Shen-Yu:TSP18} as in Sec. \ref{sub:orthogonal-TDMA-access}. 
Based on \cite[Cor. 1]{Shen-Yu:TSP18}, we replace the constraints (\ref{eq:problem-NOMA-latency-non-convex}) with stricter constraints (\ref{eq:rewrite-latenccy-constraint-TDMA-1})-(\ref{eq:rewrite-latenccy-constraint-TDMA-4}), which become equivalent to (\ref{eq:problem-NOMA-latency-non-convex}) if the variables $\lambda_{F,k}^{\text{ul}}$, $\lambda_{F,k}^{\text{dl}}$, $\lambda_{E,i_k,k}^{\text{exe}}$ and $\lambda_{C,k}^{\text{exe}}$ equal (\ref{eq:update-lambda-TDMA}).


The other non-convex constraints (\ref{eq:problem-NOMA-rate-edge-uplink}) and (\ref{eq:problem-NOMA-rate-edge-downlink}) contain ratios of matrix variables. Thus, we need to employ matrix FP \cite{Shen:TN19}, which generalizes scalar or vector version of FP in \cite{Shen-Yu:TSP18}.
From \cite[Cor. 1]{Shen:TN19}, the following constraints are stricter than  (\ref{eq:problem-NOMA-rate-edge-uplink}) and (\ref{eq:problem-NOMA-rate-edge-downlink}) for any $\Gamma_{E,k}^{\text{ul}}\in\mathbb{C}^{1\times 1}$, $\boldsymbol{\theta}_{E,k}^{\text{ul}}\in\mathbb{C}^{n_{E,i_k}\times 1}$, $\boldsymbol{\Gamma}_{E,k}^{\text{dl}}\in \mathbb{C}^{n_{E,i_k}\times n_{E,i_k}}$ and $\boldsymbol{\theta}_{E,k}^{\text{dl}}\in\mathbb{C}^{1\times n_{E,i_k}}$:
\begin{subequations} \label{eq:rewrite-constraint-edge-rate-NOMA}
\begin{align}
    R_{E,k}^{\text{ul}} &\leq \phi\left( \begin{array}{c} \Gamma_{E,k}^{\text{ul}}, \boldsymbol{\theta}_{E,k}^{\text{ul}}, \tilde{p}_k^{\text{ul}} \mathbf{h}_{i_k,k}^{\text{ul}}\boldsymbol{,} \\ \sigma_{z,\text{ul}}^2\mathbf{I} + \sum_{l\in\mathcal{N}_U} p_l^{\text{ul}} \mathbf{h}_{i_k,l}^{\text{ul}} \mathbf{h}_{i_k,l}^{\text{ul}\, H} \end{array} \right), \text{ and } \label{eq:rewrite-constraint-edge-rate-NOMA-uplink} \\
    R_{E,k}^{\text{dl}} &\leq \phi\left( \begin{array}{c} \boldsymbol{\Gamma}_{E,k}^{\text{dl}}, \boldsymbol{\theta}_{E,k}^{\text{dl}}, \mathbf{h}_{k,i_k}^{\text{dl}\,H} \tilde{\mathbf{Q}}_{E,k}^{\text{dl}}\boldsymbol{,} \\ \sigma_{z,\text{dl}}^2 + \sum_{l\in\mathcal{N}_U} \mathbf{h}_{k,i_l}^{\text{dl}\,H} \mathbf{Q}_{E,l}^{\text{dl}} \mathbf{h}_{k,i_l}^{\text{dl}} \end{array}   \right), \label{eq:rewrite-constraint-edge-rate-NOMA-downlink}
\end{align}
\end{subequations}
where we have defined the variables $\tilde{p}_k^{\text{ul}} = \sqrt{ p_k^{\text{ul}} } \in [0,\sqrt{P^{\text{ul}}}]$, $\tilde{\mathbf{Q}}_{E,k}^{\text{dl}} = \mathbf{Q}_{E,k}^{\text{dl}\,1/2}$, and the function
\begin{align}
    \phi\left( \mathbf{A}, \mathbf{B}, \mathbf{C}, \mathbf{D}\right) & = \log_2\det\left(\mathbf{I}+\mathbf{A}\right) - \frac{1}{\ln 2} \text{tr}\left(\mathbf{A}\right) \\
    & + \frac{1}{\ln 2} \text{tr}\left( \left(\mathbf{I}+\mathbf{A}\right)\left( 2\mathbf{C}^H\mathbf{B} - \mathbf{B}^H\mathbf{D}\mathbf{B} \right) \right). \nonumber
\end{align}
Also, the above constraints are equivalent to (\ref{eq:problem-NOMA-rate-edge-uplink}) and (\ref{eq:problem-NOMA-rate-edge-downlink}) if
\begin{subequations} \label{eq:update-auxiliary-NOMA}
\begin{align}
    \Gamma_{E,k}^{\text{ul}} &= p_k^{\text{ul}} \mathbf{h}_{i_k,k}^{\text{ul}\,H} \left( \! \sigma_{z,\text{ul}}^2\mathbf{I} + \!\!\!\! \sum_{l\in\mathcal{N}_U\setminus\{k\}} \!\!\!\! p_l^{\text{ul}} \mathbf{h}_{i_k,l}^{\text{ul}} \mathbf{h}_{i_k,l}^{\text{ul}\,H} \! \right)^{\!\!\! -1} \!\!\!\mathbf{h}_{i_k,k}^{\text{ul}}, \label{eq:update-gamma-NOMA-uplink} \\
    \boldsymbol{\theta}_{E,k}^{\text{ul}} &= \tilde{p}_k^{\text{ul}} \left( \sigma_{z,\text{ul}}^2\mathbf{I} + \sum_{l\in\mathcal{N}_U} p_l^{\text{ul}} \mathbf{h}_{i_k,l}^{\text{ul}} \mathbf{h}_{i_k,l}^{\text{ul}\,H} \right)^{-1}\mathbf{h}_{i_k,k}^{\text{ul}}, \label{eq:update-theta-NOMA-uplink} \\
    \boldsymbol{\Gamma}_{E,k}^{\text{dl}} &= \tilde{\mathbf{Q}}_{E,k}^{\text{dl}\,H} \mathbf{h}_{k,i_k}^{\text{dl}} \left( \sigma_{z,\text{dl}}^2 + \!\!\sum_{l\in\mathcal{N}_U\setminus\{k\}} \!\! \mathbf{h}_{k,i_l}^{\text{dl}\,H} \mathbf{Q}_{E,l}^{\text{dl}} \mathbf{h}_{k,i_l}^{\text{dl}} \right)^{-1} \nonumber \\ & \times \mathbf{h}_{k,i_k}^{\text{dl}\, H} \tilde{\mathbf{Q}}_{E,k}^{\text{dl}}, \text{ and } \label{eq:update-gamma-NOMA-downlink} \\
    \boldsymbol{\theta}_{E,k}^{\text{dl}} &= \left( \! \sigma_{z,\text{dl}}^2 + \! \sum_{l\in\mathcal{N}_U} \mathbf{h}_{k,i_l}^{\text{dl}\,H} \mathbf{Q}_{E,l}^{\text{dl}} \mathbf{h}_{k,i_l}^{\text{dl}}  \! \right)^{-1} \!\! \mathbf{h}_{k,i_k}^{\text{dl}\, H} \tilde{\mathbf{Q}}_{E,k}^{\text{dl}}. \label{eq:update-theta-NOMA-downlink}
\end{align}
\end{subequations}

Using the alternative representations (\ref{eq:rewrite-latency-constraint-TDMA}) and (\ref{eq:rewrite-constraint-edge-rate-NOMA}) to the non-convex constraints (\ref{eq:problem-NOMA-latency-non-convex})-(\ref{eq:problem-NOMA-rate-edge-downlink}), we restate the problem (\ref{eq:problem-NOMA}) with additional optimization variables $\boldsymbol{\lambda}$, $\boldsymbol{\Gamma} \triangleq \{ \Gamma_{E,k}^{\text{ul}}, \boldsymbol{\Gamma}_{E,k}^{\text{dl}} \}_{k\in\mathcal{N}_U}$ and $\boldsymbol{\theta} \triangleq \{ \boldsymbol{\theta}_{E,k}^{\text{ul}}, \boldsymbol{\theta}_{E,k}^{\text{dl}} \}_{k\in\mathcal{N}_U}$.
We tackle the obtained problem by alternately optimizing the variables $\{\mathbf{c}, \tilde{\mathbf{p}}, \tilde{\mathbf{Q}}, \mathbf{F}, \mathbf{C}_F, \boldsymbol{\tau}, \mathbf{R}\}$ and $\{\boldsymbol{\lambda}, \boldsymbol{\Gamma}, \boldsymbol{\theta} \}$. The detailed algorithm is summarized in Algorithm 2.
Similarly to Algorithm 1, Algorithm 2 achieves monotonically decreasing latency with respect to the number of iterations, whose solution converges to a locally optimal point of (\ref{eq:problem-NOMA}) due to its non-convexity.
In Sec. \ref{sec:numerical}, we initialize the variables $\{ \mathbf{c}, \mathbf{F}, \mathbf{C}_F \}$ and $\{\tilde{\mathbf{p}}, \tilde{\mathbf{Q}}\}$ as (\ref{eq:init-ck})-(\ref{eq:init-CF}) and
\begin{subequations}
\begin{align}
    &\tilde{p}_k \leftarrow \sqrt{P^{\text{ul}}}, \, k\in\mathcal{N}_U, \label{eq:init-p-ul-NOMA} \\
    &\tilde{\mathbf{Q}}_k^{\text{dl}} \leftarrow \sqrt{ \frac{P^{\text{dl}}}{ \sum_{l\in\mathcal{N}_{U,i}} || \mathbf{V}_l^{\text{dl}} ||^2_F } } \mathbf{V}_k^{\text{dl}}, \, k\in\mathcal{N}_{U,i}, i\in\mathcal{N}_E, \label{eq:init-Q-tilde-dl-NOMA}
\end{align}
\end{subequations}
respectively, where the elements of $\mathbf{V}_k^{\text{dl}}\in\mathbb{C}^{n_{E,i}\times n_{E,i}}$, $k\in\mathcal{N}_{U,i}$, are independent and identically distributed as $\mathcal{CN}(0,1)$.
For the given $\{ \mathbf{c}, \mathbf{F}, \mathbf{C}_F, \tilde{\mathbf{p}}, \tilde{\mathbf{Q}} \}$, we compute
the rates $\mathbf{R}$ using (\ref{eq:rate-edge-uplink-NOMA}) and (\ref{eq:rate-edge-downlink-NOMA}), from which the latency variables $\boldsymbol{\tau}$ can be initialized as (\ref{eq:edge-latency-uplink-TDMA}), (\ref{eq:latency-fronthaul-uplink-TDMA}), (\ref{eq:latency-fronthaul-downlink-TDMA}), and (\ref{eq:edge-latency-downlink-TDMA}).

The complexity of Algorithm 2 is given as the product of the number of iterations and the complexity of solving the convex problem at Step 4. 
The complexity of the latter is upper bounded by $\mathcal{O}( n(n^3 + M) \log(1/\epsilon))$ \cite[p. 4]{BTal-Nemirovski}, where the numbers of optimization variables and arithmetic operations are given as $n = N_U (4 \tilde{n}_E^2 + 14)$ and $M= N_U (\tilde{n}_E(14\tilde{n}_E + 1) + 41) + \tilde{n}_E (8\tilde{n}_E^2 + 5\tilde{n}_E + 3)$, respectively.
Here we have assumed that every EN uses the same number $\tilde{n}_E$ of antennas, i.e., $n_{E,i} = \tilde{n}_E$ for all $i\in\mathcal{N}_E$.
Some numerical evidence of the convergence rate of Algorithm 2 is provided in Sec. \ref{sec:numerical}.

\begin{algorithm}
\caption{Alternating optimization algorithm that tackles problem  (\ref{eq:problem-NOMA})}

\textbf{1.} Initialize $\{\mathbf{c}, \tilde{\mathbf{p}}, \tilde{\mathbf{Q}}, \mathbf{F}, \mathbf{C}_F, \boldsymbol{\tau}, \mathbf{R}\}$ as arbitrary values/matrices that satisfy the constraints (\ref{eq:problem-NOMA-latency-edge-uplink})-(\ref{eq:problem-NOMA-sum-F-C}), and set $t\leftarrow 1$.

\textbf{2.} Calculate the total latency $\tau_T$ in (\ref{eq:total-latency-DRAN}) with the initialized $\{\mathbf{c}, \tilde{\mathbf{p}}, \tilde{\mathbf{Q}}, \mathbf{F}, \mathbf{C}_F, \boldsymbol{\tau}, \mathbf{R}\}$, and set $\tau_T^{(0)}\leftarrow \tau_T$.

\textbf{3.} Set $\{\boldsymbol{\lambda},\boldsymbol{\Gamma},\boldsymbol{\theta}\}$ according to (\ref{eq:update-lambda-TDMA}) and (\ref{eq:update-auxiliary-NOMA}). 

\textbf{4.} Update the variables $\{\mathbf{c}, \tilde{\mathbf{p}}, \tilde{\mathbf{Q}}, \mathbf{F}, \mathbf{C}_F, \boldsymbol{\tau}, \mathbf{R}\}$ as a solution of the convex problem which is obtained by replacing the constraints (\ref{eq:problem-NOMA-latency-non-convex})-(\ref{eq:problem-NOMA-rate-edge-downlink}) with (\ref{eq:rewrite-latenccy-constraint-TDMA-1})-(\ref{eq:rewrite-latenccy-constraint-TDMA-4}) in (\ref{eq:problem-TDMA}), (\ref{eq:rewrite-constraint-edge-rate-NOMA-uplink}) and (\ref{eq:rewrite-constraint-edge-rate-NOMA-downlink}) and then by fixing $\{\boldsymbol{\lambda},\boldsymbol{\Gamma},\boldsymbol{\theta}\}$.

\textbf{5.} Calculate the total latency $\tau_T$ in (\ref{eq:total-latency-DRAN}) with the updated $\{\mathbf{c}, \tilde{\mathbf{p}}, \tilde{\mathbf{Q}}, \mathbf{F}, \mathbf{C}_F, \boldsymbol{\tau}, \mathbf{R}\}$, and set $\tau_T^{(t)}\leftarrow \tau_T$.

\textbf{6.} Stop if $| \tau_T^{(t)} - \tau_T^{(t-1)} |\leq \delta$ or $t>t_{\max}$. Otherwise, set $t\leftarrow t+1$ and go back to Step 2.

\end{algorithm}

\section{Optimization for the C-RAN Architecture} \label{sec:C-RAN}

In this section, we investigate the design of collaborative cloud and edge mobile computing system within a C-RAN architecture \cite{Park-et-al:SPM, Zhou-Yu:TSP16, Park-et-al:TWC}. In C-RAN, the baseband signals of distributed ENs are processed by the CP in a centralized manner for the purpose of effective interference management. 
In the following subsections, we describe the uplink and downlink communication phases and the total end-to-end latency required for completing all the tasks, and discuss the joint optimization of C-RAN signal processing and computational resource allocation strategies.

\subsection{Uplink Communication and Latency}

As illustrated in Sec. \ref{sub:computation-model-task-split}, each UE $k$ splits its computation input information into two parts of $c_k b_{I,k}$ and $(1-c_k) b_{I,k}$ bits, and sends the former and latter parts to its serving EN $i_k$ and the CP, respectively. In the D-RAN protocol detailed in Sec. \ref{sec:D-RAN}, both parts were encoded into a single codeword, since all the input information had to be decoded by the serving EN $i_k$. However, in the C-RAN scheme, only one part is decoded by EN $i_k$, and the other codeword is decoded by the CP based on the fronthaul received signals.
To accommodate this requirement, we leverage superposition coding as discussed next.

We denote the encoded signals for the two parts of $c_k b_{I,k}$ and $(1-c_k)b_{I,k}$ bits by $s_{E,k}^{\text{ul}}$ and $s_{C,k}^{\text{ul}}$, respectively. Under independent Gaussian channel codebooks, the two signals are independent of each other and distributed as $s_{E,k}^{\text{ul}}\sim \mathcal{CN}(0,p_{E,k}^{\text{ul}})$ and $s_{C,k}^{\text{ul}}\sim \mathcal{CN}(0,p_{C,k}^{\text{ul}})$. 
UE $k$ transmits a superposition of the encoded signals so that the transmit signal $x_k^{\text{ul}}$ is given as
\begin{align}
    x_k^{\text{ul}} = s_{E,k}^{\text{ul}} + s_{C,k}^{\text{ul}}, \label{eq:superposition-uplink}
\end{align}
and the transmit power constraint (\ref{eq:power-constraint-ul}) can be written as $p_{E,k}^{\text{ul}} + p_{C,k}^{\text{ul}} \leq P^{\text{ul}}$.

Based on the uplink received signal $\mathbf{y}_i^{\text{ul}}$, EN $i$ detects the signals $s_{E,k}^{\text{ul}}$ transmitted by its serving UEs $k\in\mathcal{N}_{U,i}$. The achievable rate $R_{E,k}^{\text{ul}}$ of each signal $s_{E,k}^{\text{ul}}$ in bps is given as $R_{E,k}^{\text{ul}} = W^{\text{ul}} I( s_{E,k}^{\text{ul}} ; \mathbf{y}_{i_k}^{\text{ul}} )$ with 
\begin{align}
    & I\left( s_{E,k}^{\text{ul}} ; \mathbf{y}_{i_k}^{\text{ul}} \right) = f_{E,k}^{\text{ul}}\left( \mathbf{p}^{\text{ul}} \right) = \label{eq:achievable-rate-MEC-ul} \\
    & \Psi \left( p_{E,k}^{\text{ul}}\mathbf{h}_{i_k,k}^{\text{ul}} \mathbf{h}_{i_k,k}^{\text{ul}H}\boldsymbol{,} \left(\begin{array}{c} \sigma_{z,\text{ul}}^2\mathbf{I} + \!\! \sum_{l\in\mathcal{N}_U\setminus\{k\}} p_{E,l}^{\text{ul}}\mathbf{h}_{i_k,l}^{\text{ul}} \mathbf{h}_{i_k,l}^{\text{ul} \, H} \\ + \sum_{l\in\mathcal{N}_U} p_{C,l}^{\text{ul}} \mathbf{h}_{i_k,l}^{\text{ul}}\mathbf{h}_{i_k,l}^{\text{ul}\,H} \end{array}\right)  \right). \nonumber
\end{align}
Here we have defined $\mathbf{p}^{\text{ul}} \triangleq \{p_{E,k}^{\text{ul}}, p_{C,k}^{\text{ul}}\}_{k\in\mathcal{N}_U}$.

After the local decoding described above, EN $i$ cancels out the impact of the decoded signals from the received signal $\mathbf{y}_i^{\text{ul}}$ as
\begin{align}
    \tilde{\mathbf{y}}_i^{\text{ul}} \leftarrow \mathbf{y}_i^{\text{ul}} - \sum\nolimits_{l \in \mathcal{N}_{U,i}} \mathbf{h}_{i,l}^{\text{ul}} s_{E,l}^{\text{ul}}. \label{eq:local-cancellation}
\end{align}
Since the fronthaul link connecting EN $i$ to the CP has finite capacity $C_F$ bps, a quantized version of the signal $\tilde{\mathbf{y}}_i^{\text{ul}}$, denoted by $\hat{\mathbf{y}}_i^{\text{ul}}$, is forwarded to the CP. We assume the Gaussian test channel as in \cite{Zhou-Yu:TSP16, Park-et-al:TWC}. Then, the quantized signal $\hat{\mathbf{y}}_i^{\text{ul}}$ is modeled as
\begin{align}
    \hat{\mathbf{y}}_i^{\text{ul}} = \tilde{\mathbf{y}}_i^{\text{ul}} + \mathbf{q}_i^{\text{ul}}, \label{eq:Gaussian-test-channel-ul}
\end{align}
where the quantization distortion noise $\mathbf{q}_i^{\text{ul}}$ is independent of $\tilde{\mathbf{y}}_i^{\text{ul}}$ and is distributed as $\mathbf{q}_i^{\text{ul}}\sim\mathcal{CN}(\mathbf{0}, \mathbf{\Omega}_i^{\text{ul}})$.
Under the quantization model (\ref{eq:Gaussian-test-channel-ul}), the compression rate $\gamma_i^{\text{ul}}$, that equals the number of bits representing the quantized signal $\hat{\mathbf{y}}_i^{\text{ul}}$ per baseband sample, is given as \cite{ElGamal-Kim}
\begin{align}
    & \gamma_i^{\text{ul}} = I\left( \tilde{\mathbf{y}}_i^{\text{ul}}; \hat{\mathbf{y}}_i^{\text{ul}} \right) = g_i^{\text{ul}}\left( \mathbf{p}^{\text{ul}}, \mathbf{\Omega}_i^{\text{ul}} \right)  \nonumber \\
    &= \log_2\det\left( \begin{array}{c} \sum_{k\in\mathcal{N}_U \setminus \mathcal{N}_{U,i}} p_{E,k}^{\text{ul}} \mathbf{h}_{i,k}^{\text{ul}}  \mathbf{h}_{i,k}^{\text{ul}\,H} + \\
    \sum_{k\in\mathcal{N}_U} p_{C,k}^{\text{ul}} \mathbf{h}_{i,k}^{\text{ul}}  \mathbf{h}_{i,k}^{\text{ul}\,H} + \sigma_{z,\text{ul}}^2\mathbf{I} + \mathbf{\Omega}_i^{\text{ul}}
    \end{array} \right) \nonumber \\
    & - \log_2\det\left( \mathbf{\Omega}_i^{\text{ul}} \right). \label{eq:compression-rate-ul}
\end{align}

EN $i$ should send $W^{\text{ul}} \tau_E^{\text{ul}} \gamma_i^{\text{ul}}$ bits to the CP on the fronthaul link of capacity $C_F$ bps, since the duration of each baseband sample is approximately $1/W^{\text{ul}}$ sec, and hence $\tau_E^{\text{ul}} / (1/W^{\text{ul}}) = W^{\text{ul}} \tau_E^{\text{ul}}$ quanzited baseband samples should be forwarded to the CP. 
Due to the parallel operation of fronthaul links of different ENs, the fronthaul latency $\tau_F^{\text{ul}}$ for uplink is given as
\begin{align}
    \tau_F^{\text{ul}} = \max_{i\in\mathcal{N}_E} \frac{W^{\text{ul}} \tau_E^{\text{ul}} \gamma_i^{\text{ul}}}{C_F}. \label{eq:fronthaul-latency-uplink}
\end{align}

The CP recovers the quantized signals $\hat{\mathbf{y}}_1^{\text{ul}},\hat{\mathbf{y}}_2^{\text{ul}},\ldots,\hat{\mathbf{y}}_{N_E}^{\text{ul}}$ from the bit streams received on the fronthaul links. 
The vector $\hat{\mathbf{y}}^{\text{ul}} = [\hat{\mathbf{y}}^{\text{ul}H}_1 \, \hat{\mathbf{y}}^{\text{ul}H}_2 \cdots \hat{\mathbf{y}}^{\text{ul}H}_{N_E}]^H$, which stacks the quantized signals from all ENs, can be written as
\begin{align}
    \hat{\mathbf{y}}^{\text{ul}} = \sum\nolimits_{l\in\mathcal{N}_U} \tilde{\mathbf{h}}_l^{\text{ul}} s_{E,l}^{\text{ul}} + \sum\nolimits_{l\in\mathcal{N}_U} \mathbf{h}_l^{\text{ul}} s_{C,l}^{\text{ul}} + \mathbf{q}^{\text{ul}} + \mathbf{z}^{\text{ul}}, \label{eq:stacked-quantized-signals}
\end{align}
where we have defined $\mathbf{h}_k^{\text{ul}} = [\mathbf{h}_{1,k}^{\text{ul}H} \, \mathbf{h}_{2,k}^{\text{ul}H} \cdots \mathbf{h}_{N_E,k}^{\text{ul}H}]^H$, $\tilde{\mathbf{h}}_k^{\text{ul}} = [\tilde{\mathbf{h}}_{1,k}^{\text{ul}H} \, \tilde{\mathbf{h}}_{2,k}^{\text{ul}H} \cdots \tilde{\mathbf{h}}_{N_E,k}^{\text{ul}H}]^H$ with $\tilde{\mathbf{h}}_{i,k}^{\text{ul}} = \mathbf{h}_{i,k} \mathbf{1}_{i\neq i_k} + \mathbf{0}_{n_{E,i} \times 1} \mathbf{1}_{i = i_k}$, $\mathbf{q}^{\text{ul}} = [\mathbf{q}_1^{\text{ul}H} \, \mathbf{q}_2^{\text{ul}H} \cdots \mathbf{q}_{N_E}^{\text{ul}H}]^H$, and $\mathbf{z}^{\text{ul}} = [\mathbf{z}_1^{\text{ul}H} \, \mathbf{z}_2^{\text{ul}H} \cdots \mathbf{z}_{N_E}^{\text{ul}H}]^H$. 
Here $\mathbf{1}_{(\cdot)}$ is an indicator function which takes 1 if the statement in the subscript is true and 0 otherwise. The stacked noise vectors $\mathbf{q}^{\text{ul}}$ and $\mathbf{z}^{\text{ul}}$ are distributed as $\mathbf{q}^{\text{ul}}\sim\mathcal{CN}(\mathbf{0}, \bar{\mathbf{\Omega}}^{\text{ul}})$ and $\mathbf{z}^{\text{ul}}\sim\mathcal{CN}(\mathbf{0}, \sigma_{z,\text{ul}}^2\mathbf{I})$, respectively, with $\bar{\mathbf{\Omega}}^{\text{ul}} = \text{diag}(\{\mathbf{\Omega}^{\text{ul}}_i\}_{i\in\mathcal{N}_E})$.

Using the recovered quantized signal vector $\hat{\mathbf{y}}^{\text{ul}}$, the CP detects all the signals $s_{C,k}^{\text{ul}}$, which are necessary for cloud computing. The achievable rate $R_{C,k}^{\text{ul}}$ of the signal $s_{C,k}^{\text{ul}}$ is given as $R_{C,k}^{\text{ul}} = W^{\text{ul}} I( s_{C,k}^{\text{ul}}; \hat{\mathbf{y}}^{\text{ul}} )$, where the mutual information value is computed as
\begin{align}
     & I\left( s_{C,k}^{\text{ul}}; \hat{\mathbf{y}}^{\text{ul}} \right) = f_{C,k}^{\text{ul}}\left( \mathbf{p}^{\text{ul}}, \mathbf{\Omega}^{\text{ul}} \right) = \label{eq:achievable-rate-MCC-ul} \\
    & \Psi\left( p_{C,k}^{\text{ul}} \mathbf{h}_k^{\text{ul}} \mathbf{h}_k^{\text{ul}H}\boldsymbol{,} \left( \begin{array}{c}  \sigma_{z,\text{ul}}^2\mathbf{I} + \sum_{l\in\mathcal{N}_U} p_{E,l}^{\text{ul}} \tilde{\mathbf{h}}_l^{\text{ul}}  \tilde{\mathbf{h}}_l^{\text{ul}H} +\\ \sum_{l\in\mathcal{N}_U\setminus \{k\}} p_{C,l}^{\text{ul}} \mathbf{h}_l^{\text{ul}}  \mathbf{h}_l^{\text{ul}H} + \bar{\mathbf{\Omega}}^{\text{ul}}  \end{array}  \right) \right). \nonumber
\end{align}

Consequently, the latency $\tau_{E}^{\text{ul}}$ for uploading the input information of the UEs on the uplink channel is given as
\begin{align}
    \tau_{E}^{\text{ul}} = \max_{k\in\mathcal{N}_U} \bigg\{ \frac{c_k b_{I,k}}{ W^{\text{ul}}f_{E,k}^{\text{ul}}\left(\mathbf{p}\right) } , \frac{(1-c_k) b_{I,k}}{W^{\text{ul}} f_{C,k}^{\text{ul}}\left(\mathbf{p}, \mathbf{\Omega}^{\text{ul}}\right)} \bigg\}. \label{eq:latency-edge-MEC-uplink}
\end{align}

\subsection{Downlink Communication and Latency}

After completing the computation tasks, the CP encodes the computation output information of $(1-c_k)b_{O,k}$ bits for each UE $k$ with Gaussian channel codebook and obtains an encoded baseband signal $\mathbf{s}_{C,k}^{\text{dl}} \in \mathbb{C}^{n_E\times 1} \sim\mathcal{CN}(\mathbf{0},\mathbf{Q}^{\text{dl}}_{C,k})$.

The CP computes a signal vector $\tilde{\mathbf{x}}^{\text{dl}}\in\mathbb{C}^{n_E\times 1}$ by superimposing the encoded signals as
\begin{align}
    \tilde{\mathbf{x}}^{\text{dl}} = \sum\nolimits_{k\in\mathcal{N}_{U}} \mathbf{s}_{C,k}^{\text{dl}}. \label{eq:superposition-CP-dl}
\end{align}
The $i$th subvector $\tilde{\mathbf{x}}_i^{\text{dl}}\in\mathbb{C}^{n_{E,i}\times 1}$ of $\tilde{\mathbf{x}}^{\text{dl}} = [\tilde{\mathbf{x}}_1^{\text{dl}H} \cdots \tilde{\mathbf{x}}_{N_E}^{\text{dl}H}]^H$ is transferred to EN $i$ on the fronthaul link. To this end, it is quantized, and we model the quantized signal $\hat{\mathbf{x}}_i^{\text{dl}}$ under the Gaussian test channel \cite{Zhou-Yu:TSP16, Park-et-al:TWC} as
\begin{align}
    \hat{\mathbf{x}}_i^{\text{dl}} = \tilde{\mathbf{x}}_i^{\text{dl}} + \mathbf{q}_i^{\text{dl}}, \label{eq:Gaussian-test-channel-dl}
\end{align}
where the quantization distortion noise $\mathbf{q}_i^{\text{dl}}$ is independent of $\tilde{\mathbf{x}}_i^{\text{dl}}$ and distributed as $\mathbf{q}_i^{\text{dl}}\sim\mathcal{CN}(\mathbf{0}, \mathbf{\Omega}_i^{\text{dl}})$.

The compression rate $\gamma_i^{\text{dl}}$ needed for representing the quantized signal $\hat{\mathbf{x}}_i^{\text{dl}}$ in bits per baseband sample is given as
\begin{align}
    & \gamma_i^{\text{dl}} = I\left( \tilde{\mathbf{x}}_i^{\text{dl}}; \hat{\mathbf{x}}_i^{\text{dl}} \right) = g_i^{\text{dl}}\left( \mathbf{Q}^{\text{dl}}, \mathbf{\Omega}^{\text{dl}}_i \right) = \label{eq:compression-rate-dl} \\
    & \log_2\det\left( \sum\nolimits_{k\in\mathcal{N}_{U}} \mathbf{E}_i^H\mathbf{Q}_{C,k}^{\text{dl}}\mathbf{E}_i + \mathbf{\Omega}_i^{\text{dl}} \right) - \log_2\det\left( \mathbf{\Omega}_i^{\text{dl}} \right), \nonumber
\end{align}
where the elements of $\mathbf{E}_i\in\mathbb{C}^{n_E\times n_{E,i}}$ are filled with zeros except for the rows from $\sum_{j=1}^{i-1}n_{E,j}+1$ to $\sum_{j=1}^i n_{E,j}$ being an identity matrix of size $n_{E,i}\times n_{E,i}$.

Similar to (\ref{eq:fronthaul-latency-uplink}) for uplink, the downlink fronthaul latency $\tau_F^{\text{dl}}$ for given $\gamma_i^{\text{dl}}$, $i\in\mathcal{N}_E$, and $\tau_E^{\text{dl}}$ is computed as
\begin{align}
    \tau_F^{\text{dl}} = \max_{i\in\mathcal{N}_E} \frac{W^{\text{dl}} \tau_E^{\text{dl}} \gamma_i^{\text{dl}} }{ C_F }. \label{eq:fronthaul-latency-dl}
\end{align}

Each EN $i$ also encodes the edge computation output information for UE $k\in\mathcal{N}_{U,i}$ of $c_k b_{O,k}$ bits producing an encoded baseband signal $\mathbf{s}_{E,k}^{\text{dl}}\in\mathbb{C}^{n_{E,i}\times 1}\sim\mathcal{CN}(\mathbf{0}, \mathbf{Q}_{E,k}^{\text{dl}})$.
EN $i$ then transmits a superposition of the locally encoded signals $\mathbf{s}_{E,k}^{\text{dl}}$, $k\in\mathcal{N}_{U,i}$, and the quantized signal $\hat{\mathbf{x}}_i^{\text{dl}}$, which was received on fronthaul, over the downlink channel to UEs. Thus, the signal $\mathbf{x}_i^{\text{dl}}$ transmitted by EN $i$ is given as
\begin{align}
    \mathbf{x}_i^{\text{dl}} = \sum\nolimits_{k\in\mathcal{N}_{U,i}}  \mathbf{s}_{E,k}^{\text{dl}} + \hat{\mathbf{x}}_i^{\text{dl}}. \label{eq:superposition-EN-i-dl}
\end{align}
With (\ref{eq:superposition-EN-i-dl}), the transmit power constraint (\ref{eq:power-constraint-dl}) at EN $i$ can be written as
\begin{align}
    \sum_{k\in\mathcal{N}_{U,i}}\!\!\! \text{tr}\left(\mathbf{Q}_{E,k}^{\text{dl}}\right) + \!\! \sum_{k\in\mathcal{N}_{U}}\!\!\!\text{tr}\left(\mathbf{E}_i^H\mathbf{Q}_{C,k}^{\text{dl}}\mathbf{E}_i\right) \! + \! \text{tr}\left( \mathbf{\Omega}_i^{\text{dl}} \right) \leq P^{\text{dl}}. \label{eq:power-constraint-dl-rewritten}
\end{align}
The first term in the left-hand side (LHS) measures the power of the signals $\{\mathbf{s}_{E,k}^{\text{dl}}\}_{k\in\mathcal{N}_{U,i}}$, which encode the computation output information processed by EN $i$. The sum of the second and third terms is the power of the signal $\hat{\mathbf{x}}_i^{\text{dl}}$, which is a quantized version of $\tilde{\mathbf{x}}_i^{\text{dl}}$ that encodes the signals $\{\mathbf{s}_{C,k}\}_{k\in\mathcal{N}_U}$ processed by the CP.

Each UE $k$ detects the signals $\mathbf{s}_{E,k}^\text{dl}$ and $\mathbf{s}_{C,k}^\text{dl}$ based on the downlink received signal $y_k^{\text{dl}}$. 
The achievable rates of $\mathbf{s}_{E,k}^\text{dl}$ and $\mathbf{s}_{C,k}^\text{dl}$ are given as $R_{E,k}^{\text{dl}} = W^{\text{dl}} I( \mathbf{s}_{E,k}^{\text{dl}} ; y_k^{\text{dl}} )$ and $R_{C,k}^{\text{dl}} = W^{\text{dl}} I( \mathbf{s}_{C,k}^{\text{dl}} ; y_k^{\text{dl}} )$, respectively, with
\begin{subequations} \label{eq:achievable-rate-dl}
\begin{align}
    & I\left( \mathbf{s}_{E,k}^{\text{dl}} ; y_k^{\text{dl}} \right) = f_{E,k}^{\text{dl}}\left( \mathbf{Q}^{\text{dl}}, \mathbf{\Omega}^{\text{dl}} \right) = \label{eq:achievable-rate-edge-dl} \\
    & \Psi\left( \mathbf{h}_{k,i_k}^{\text{dl}H} \mathbf{Q}_{E,k}^{\text{dl}} \mathbf{h}_{k,i_k}^{\text{dl}}\boldsymbol{,} \left( \begin{array}{c} \sum_{l\in \mathcal{N}_{U} \setminus\{k\}} \mathbf{h}_{k,i_l}^{\text{dl}H}\mathbf{Q}_{E,l}^{\text{dl}}\mathbf{h}_{k,i_l}^{\text{dl}} \\
    + \sum_{l\in\mathcal{N}_U} \mathbf{h}_k^{\text{dl}H} \mathbf{Q}_{C,l}^{\text{dl}} \mathbf{h}_k^{\text{dl}} \\
     + \sigma_{z,\text{dl}}^2 + \mathbf{h}_k^{\text{dl}H} \bar{\mathbf{\Omega}}^{\text{dl}} \mathbf{h}_k^{\text{dl}} \end{array}   \right) \right), \text{ and } \nonumber \\
     & I\left( \mathbf{s}_{C,k}^{\text{dl}} ; y_k^{\text{dl}} \right) = f_{C,k}^{\text{dl}}\left( \mathbf{Q}^{\text{dl}}, \mathbf{\Omega}^{\text{dl}} \right) = \label{eq:achievable-rate-MCC-dl} \\
     & \Psi\left( \mathbf{h}_{k}^{\text{dl}H} \mathbf{Q}_{C,k}^{\text{dl}} \mathbf{h}_{k}^{\text{dl}}\boldsymbol{,} \left( \begin{array}{c} \sum_{l\in \mathcal{N}_{U} } \mathbf{h}_{k,i_l}^{\text{dl}H}\mathbf{Q}_{E,l}^{\text{dl}}\mathbf{h}_{k,i_l}^{\text{dl}} + \\ \sum_{l\in\mathcal{N}_U \setminus \{k\}} \mathbf{h}_k^{\text{dl}H} \mathbf{Q}_{C,l}^{\text{dl}} \mathbf{h}_k^{\text{dl}} \\
     + \sigma_{z,\text{dl}}^2 + \mathbf{h}_k^{\text{dl}H} \bar{\mathbf{\Omega}}^{\text{dl}} \mathbf{h}_k^{\text{dl}} \end{array}   \right) \right). \nonumber
\end{align}
\end{subequations}
Here, we have defined $\mathbf{h}_k^{\text{dl}} = [ \mathbf{h}_{k,1}^{\text{dl}H} \, \mathbf{h}_{k,2}^{\text{dl}H} \cdots \mathbf{h}_{k,N_E}^{\text{dl}H} ]^H$ and $\bar{\mathbf{\Omega}}^{\text{dl}} = \text{diag}(\{ \mathbf{\Omega}_i^{\text{dl}} \}_{i\in\mathcal{N}_E})$.

With the downlink rates described above, the latency $\tau_{E}^{\text{dl}}$ for downloading the output information on the downlink channel is given as
\begin{align}
    \tau_{E}^{\text{dl}} = \max_{k\in\mathcal{N}_U}\!\! \bigg\{ \frac{c_k b_{O,k}}{W^{\text{dl}} f_{E,k}^{\text{dl}}\!\left( \mathbf{Q}^{\text{dl}}, \mathbf{\Omega}^{\text{dl}} \right) } ,  \frac{(1-c_k) b_{O,k}}{W^{\text{dl}} f_{C,k}^{\text{dl}}\!\left( \mathbf{Q}^{\text{dl}}, \mathbf{\Omega}^{\text{dl}} \right) } \bigg\}. \label{eq:latency-edge-MEC-downlink}
\end{align}

\subsection{Total End-to-End Latency With C-RAN}

The total end-to-end latency $\tau_T$ for completing the all the tasks within the described C-RAN architecture is modeled as
\begin{align}
    \tau_T = \tau_E^{\text{ul}} + \max\Big\{ \tau_E^{\text{exe}}, \tau_F^{\text{ul}} + \tau_C^{\text{exe}} + \tau_F^{\text{dl}} \Big\} + \tau_E^{\text{dl}}, \label{eq:total-latency}
\end{align}
where the fronthaul latency $\tau_{F}^{\text{ul}}$, $\tau_{F}^{\text{dl}}$ and the edge latency $\tau_{E}^{\text{ul}}$, $\tau_{E}^{\text{dl}}$ are defined in (\ref{eq:fronthaul-latency-uplink}), (\ref{eq:fronthaul-latency-dl}), (\ref{eq:latency-edge-MEC-uplink}) and (\ref{eq:latency-edge-MEC-downlink}), respectively. Also, $\tau_E^{\text{exe}}$ and $\tau_C^{\text{exe}}$ represent the latency for executing the computation tasks at the ENs and CP which are are given as
\begin{align}
    \tau_E^{\text{exe}} = \max_{k\in\mathcal{N}_U} \tau_{E,i_k,k}^{\text{exe}} \,\,\,\,\,\, \text{and} \,\,\,\,\,
    \tau_C^{\text{exe}} = \max_{k\in\mathcal{N}_U} \tau_{C,k}^{\text{exe}}, \label{eq:computing-latency-CRAN}
\end{align}
with $\tau_{E,i_k,k}^{\text{exe}}$ and $\tau_{C,k}^{\text{exe}}$ in (\ref{eq:edge-computing-latency-EN-i}) and (\ref{eq:cloud-computing-latency-CP}).

\subsection{Optimization} \label{sub:optimization}

We aim at jointly optimizing the task splitting variables $\mathbf{c}$, the uplink $\{\mathbf{p}^{\text{ul}}, \mathbf{\Omega}^{\text{ul}}\}$ and downlink communication strategies $\{\mathbf{Q}^{\text{dl}}, \mathbf{\Omega}^{\text{dl}}\}$ with the goal of minimizing the end-to-end latency $\tau_T$ in (\ref{eq:total-latency}). The problem at hand can be stated as
\begin{subequations}\label{eq:problem-original}
\begin{align}
\underset{ ^{\mathbf{p}\geq 0, \mathbf{c}\geq 0, \mathbf{Q}\succeq\mathbf{0},} _{\,\,\,\,\,\,\mathbf{\Omega}\succeq\mathbf{0},\mathbf{F}, \boldsymbol{\tau},\mathbf{R}} }{\mathrm{minimize}}\,\, &  \tau_E^{\text{ul}} + \max\Big\{ \tau_{E}^{\text{exe}}, \tau_F^{\text{ul}} + \tau_{C}^{\text{exe}} + \tau_F^{\text{dl}} \Big\} + \tau_E^{\text{dl}} \label{eq:problem-original-cost}\\
\mathrm{s.t.}\,\,\,\,\,\,\,\,\,\, & \tau_E^{\text{ul}} \geq \frac{c_k b_{I,k}}{R_{E,k}^{\text{ul}}}, \,\, k\in\mathcal{N}_U, \label{eq:problem-original-latency-edge-ul-MEC} \\
& \tau_E^{\text{ul}} \geq \frac{(1-c_k) b_{I,k}}{R_{C,k}^{\text{ul}}},\,\,k\in\mathcal{N}_U,\label{eq:problem-original-latency-edge-ul-MCC}\\
 & \tau_F^{\text{ul}} \geq \frac{ W^{\text{ul}} \tau_E^{\text{ul}} \, g_i^{\text{ul}} \left( \mathbf{p}^{\text{ul}}, \mathbf{\Omega}_i^{\text{ul}} \right) }{C_F},\,\,i\in\mathcal{N}_E,\label{eq:problem-original-fronthaul-ul} \\
 & \tau_E^{\text{dl}} \geq  \frac{c_k b_{O,k}}{R_{E,k}^{\text{dl}}}, \,\, k\in\mathcal{N}_U, \label{eq:problem-original-latency-edge-dl-MEC} \\
 & \tau_E^{\text{dl}} \geq  \frac{(1-c_k) b_{O,k}}{R_{C,k}^{\text{dl}}}, \,\, k\in\mathcal{N}_U, \label{eq:problem-original-latency-edge-dl-MCC} \\
 & \tau_F^{\text{dl}} \geq \frac{ W^{\text{dl}} \tau_E^{\text{dl}} \, g_i^{\text{dl}} \left( \mathbf{Q}^{\text{dl}}, \mathbf{\Omega}_i^{\text{dl}} \right) }{C_F},\,\,i\in\mathcal{N}_E,\label{eq:problem-original-fronthaul-dl} \\
 & \text{(\ref{eq:problem-TDMA-latency-exe-edge}), (\ref{eq:problem-TDMA-latency-exe-cloud})}, \label{eq:problem-original-latency-execution} \\
 & R_{E,k}^{\text{ul}} \leq W^{\text{ul}} f_{E,k}^{\text{ul}}\left(\mathbf{p}^{\text{ul}}\right) ,\,\,k\in\mathcal{N}_U,\label{eq:problem-original-rate-edge-ul-MEC}\\
 & R_{C,k}^{\text{ul}} \leq W^{\text{ul}}f_{C,k}^{\text{ul}}\left(\mathbf{p}^{\text{ul}}, \mathbf{\Omega}^{\text{ul}}\right), \,\, k\in\mathcal{N}_U, \\
 & R_{E,k}^{\text{dl}} \leq W^{\text{dl}} f_{E,k}^{\text{dl}}\left( \mathbf{Q}^{\text{dl}}, \mathbf{\Omega}^{\text{dl}} \right), \,\, k\in \mathcal{N}_U \\
 & R_{C,k}^{\text{dl}} \leq W^{\text{dl}} f_{C,k}^{\text{dl}}\left( \mathbf{Q}^{\text{dl}}, \mathbf{\Omega}^{\text{dl}} \right), \,\, k\in\mathcal{N}_U, \label{eq:problem-original-rate-edge-dl-MCC} \\
 & \text{(\ref{eq:problem-TDMA-computing-allocation-edge})-(\ref{eq:problem-sum-CF-downlink})}, \label{eq:problem-original-sum-F-C} \\
 & p^{\text{ul}}_{E,k} + p^{\text{ul}}_{C,k} \leq P^{\text{ul}}, \,\, k\in\mathcal{N}_U, \label{eq:problem-original-power-ul} \\
 & \sum_{k\in\mathcal{N}_{U,i}} \text{tr}\left(\mathbf{Q}_{E,k}^{\text{dl}}\right) + \sum_{k\in\mathcal{N}_{U}}\text{tr}\left(\mathbf{E}_i^H\mathbf{Q}_{C,k}^{\text{dl}}\mathbf{E}_i\right) \nonumber \\
 & \,\,\,\,\,+ \text{tr}\left( \mathbf{\Omega}_i^{\text{dl}} \right) \leq P^{\text{dl}}, \,\, i\in\mathcal{N}_E, \label{eq:problem-original-power-dl} \\
 & c_k \in [0,1], \,\, k\in\mathcal{N}_U. \label{eq:problem-original-binary-sum}
\end{align}
\end{subequations}

We note that it is more difficult to solve problem (\ref{eq:problem-original}) than problems (\ref{eq:problem-TDMA}) and (\ref{eq:problem-NOMA}) for D-RAN, since (\ref{eq:problem-original}) involves more optimization variables including the fronthaul quantization strategies $\boldsymbol{\Omega}^{\text{ul}}$ and $\boldsymbol{\Omega}^{\text{dl}}$; and the constraints (\ref{eq:problem-original-fronthaul-ul}) and (\ref{eq:problem-original-fronthaul-dl}) on the fronthaul latency have a more complicated form than (\ref{eq:problem-TDMA-latency-fronthaul-uplink}) and (\ref{eq:problem-TDMA-latency-fronthaul-downlink}) for D-RAN systems.
To address these complications, we apply FP and matrix FP \cite{Shen-Yu:TSP18, Shen:TN19} as in the methodology outlined above for D-RAN as well as the convex approximation method introduced in \cite[Lem. 1]{Zhou-Yu:TSP16}.



To this end, we first replace the constraints  (\ref{eq:problem-original-latency-execution}) with (\ref{eq:rewrite-latency-constraint-TDMA-3}) and (\ref{eq:rewrite-latenccy-constraint-TDMA-4}) which are convex for fixed $\lambda_{E,i_k,k}^{\text{exe}}$ and $\lambda_{C,k}^{\text{exe}}$ and become equivalent to (\ref{eq:problem-original-latency-execution}) when $\lambda_{E,i_k,k}^{\text{exe}}$ and $\lambda_{C,k}^{\text{exe}}$ are given as (\ref{eq:update-lambda-TDMA}).
Similarly, based on \cite[Cor. 1]{Shen-Yu:TSP18}, we consider the following constraints which are stricter than (\ref{eq:problem-original-latency-edge-ul-MEC}), (\ref{eq:problem-original-latency-edge-ul-MCC}), (\ref{eq:problem-original-latency-edge-dl-MEC}) and (\ref{eq:problem-original-latency-edge-dl-MCC}):
\begin{subequations} \label{eq:consraints-rewritten-latency-proposed}
\begin{align}
    2\lambda_{E,k}^{\text{ul}} \sqrt{ \tau_E^{\text{ul}} } - (\lambda_{E,k}^{\text{ul}})^2 c_k & \geq \frac{b_{I,k}}{R_{E,k}^{\text{ul}}}, \,\, k\in\mathcal{N}_U, \label{eq:consraints-rewritten-latency-edge-proposed-1} \\
    2\lambda_{C,k}^{\text{ul}} \sqrt{ \tau_E^{\text{ul}} } - (\lambda_{C,k}^{\text{ul}})^2 (1-c_k) & \geq \frac{b_{I,k}}{R_{C,k}^{\text{ul}}} \,\, k\in\mathcal{N}_U, \label{eq:consraints-rewritten-latency-edge-proposed-2} \\
    2\lambda_{E,k}^{\text{dl}} \sqrt{ \tau_E^{\text{dl}} } - (\lambda_{E,k}^{\text{dl}})^2 c_k & \geq \frac{b_{O,k}}{R_{E,k}^{\text{dl}}}, \,\, k\in\mathcal{N}_U, \label{eq:consraints-rewritten-latency-edge-proposed-3} \\
    2\lambda_{C,k}^{\text{dl}} \sqrt{\tau_E^{\text{dl}}} - (\lambda_{C,k}^{\text{dl}})^2 (1-c_k) & \geq \frac{b_{O,k}}{R_{C,k}^{\text{dl}}}, \,\, k\in\mathcal{N}_U. \label{eq:consraints-rewritten-latency-edge-proposed-4}
\end{align}
\end{subequations}
The above constraints become equivalent to (\ref{eq:problem-original-latency-edge-ul-MEC}), (\ref{eq:problem-original-latency-edge-ul-MCC}), (\ref{eq:problem-original-latency-edge-dl-MEC}) and (\ref{eq:problem-original-latency-edge-dl-MCC}) if
\begin{align}
    &\lambda_{E,k}^{m} = \frac{\sqrt{\tau_E^{m}}}{c_k} \, \text{ and } \, \lambda_{C,k}^{m} = \frac{\sqrt{\tau_E^{m}}}{1-c_k}, \label{eq:update-lambda-latency-edge-proposed}
\end{align}
for $m\in\{\text{ul}, \text{dl}\}$.

Now, we discuss the non-convex constraints (\ref{eq:problem-original-fronthaul-ul}) and (\ref{eq:problem-original-fronthaul-dl}). Using the epigraph form, the constraint (\ref{eq:problem-original-fronthaul-ul}) can be restated as
\begin{subequations} \label{eq:constraint-fronthaul-ul}
\begin{align}
    \tau_F^{\text{ul}} & \geq \frac{W^{\text{ul}} \tau_E^{\text{ul}} \gamma_i^{\text{ul}} }{C_F}, \,\, i\in\mathcal{N}_E, \text{ and } \label{eq:constraint-fronthaul-ul-1} \\
    \gamma_i^{\text{ul}} & \geq g_i^{\text{ul}} \left( \mathbf{p}^{\text{ul}}, \mathbf{\Omega}^{\text{ul}} \right), \,\, i\in\mathcal{N}_E. \label{eq:constraint-fronthaul-ul-2}
\end{align}
\end{subequations}
From \cite[Cor. 1]{Shen-Yu:TSP18} and \cite[Lem. 1]{Zhou-Yu:TSP16}, the following constraints are stricter than (\ref{eq:constraint-fronthaul-ul}):
\begin{subequations} \label{eq:constraint-rewritten-fronthaul-ul}
\begin{align}
    \frac{W^{\text{ul}} \gamma_i^{\text{ul}}}{C_F} &\leq 2\alpha^{\text{ul}} \sqrt{\tau_F^{\text{ul}}} - (\alpha^{\text{ul}})^2 \tau_E^{\text{ul}}, \,\, i\in\mathcal{N}_E, \text{ and } \label{eq:constraint-rewritten-fronthaul-ul-1} \\
    \gamma_i^{\text{ul}} & \geq \log_2\det\left( \mathbf{\Sigma}_i^{\text{ul}} \right) + \frac{1}{\ln 2} \times \nonumber \\ 
    & \text{tr}\left( \mathbf{\Sigma}_i^{\text{ul}-1} \left( \begin{array}{c} \sum_{k\in\mathcal{N}_U \setminus \mathcal{N}_{U,i}} p_{E,k}^{\text{ul}} \mathbf{h}_{i,k}^{\text{ul}}  \mathbf{h}_{i,k}^{\text{ul}H} \\
    + \sum_{k\in\mathcal{N}_U} p_{C,k}^{\text{ul}} \mathbf{h}_{i,k}^{\text{ul}}  \mathbf{h}_{i,k}^{\text{ul}H}
    \\ + \sigma_{z,\text{ul}}^2\mathbf{I} + \mathbf{\Omega}_i^{\text{ul}}
    \end{array} \right) \right) \nonumber \\
    & -\frac{n_{E,i}}{\ln 2} - \log_2\det\left( \mathbf{\Omega}_i^{\text{ul}} \right), \,\, i\in\mathcal{N}_E. \label{eq:constraint-rewritten-fronthaul-ul-2}
\end{align}
\end{subequations}
If we fix the auxiliary variables $\alpha^{\text{ul}}$ and $\mathbf{\Sigma}_i^{\text{ul}}$, the constraints (\ref{eq:constraint-rewritten-fronthaul-ul}) are convex.
Also, they become equivalent to (\ref{eq:constraint-fronthaul-ul}) if the auxiliary variables $\alpha^{\text{ul}}$ and $\mathbf{\Sigma}_i^{\text{ul}}$ are given as
\begin{subequations} \label{eq:auxiliary-variable-fronthaul-ul}
\begin{align}
    \alpha^{\text{ul}} & = \frac{\sqrt{\tau_F^{\text{ul}}}}{\tau_E^{\text{ul}}}, \text{ and } \label{eq:auxiliary-variable-fronthaul-ul-1} \\
    \mathbf{\Sigma}_i^{\text{ul}} &=  \sum_{k\in\mathcal{N}_U \setminus \mathcal{N}_{U,i}} p_{E,k}^{\text{ul}} \mathbf{h}_{i,k}^{\text{ul}} \mathbf{h}_{i,k}^{\text{ul}H} +
    \sum_{k\in\mathcal{N}_U} p_{C,k}^{\text{ul}} \mathbf{h}_{i,k}^{\text{ul}}  \mathbf{h}_{i,k}^{\text{ul}H} \nonumber \\
    & + \sigma_{z,\text{ul}}^2\mathbf{I} + \mathbf{\Omega}_i^{\text{ul}}. \label{eq:auxiliary-variable-fronthaul-ul-2}
\end{align}
\end{subequations}

Similarly, instead of (\ref{eq:problem-original-fronthaul-dl}) for downlink, we consider the following stricter constraints:
\begin{subequations} \label{eq:constraint-rewritten-fronthaul-dl}
\begin{align}
    \frac{W^{\text{dl}} \gamma_i^{\text{dl}}}{C_F} &\leq 2\alpha^{\text{dl}} \sqrt{\tau_F^{\text{dl}}} - (\alpha^{\text{dl}})^2 \tau_E^{\text{dl}}, \,\, i\in\mathcal{N}_E, \text{ and } \label{eq:constraint-rewritten-fronthaul-dl-1} \\
    \gamma_i^{\text{dl}} & \geq \log_2\det\left( \mathbf{\Sigma}_i^{\text{dl}} \right) + \frac{1}{\ln 2} \times \nonumber \\
    & \text{tr}\left( \mathbf{\Sigma}_i^{\text{dl}-1} \left( \begin{array}{c} 
    \sum_{k\in\mathcal{N}_U}  \mathbf{E}_i^H \tilde{\mathbf{Q}}_{C,k}^{\text{dl}}\tilde{\mathbf{Q}}_{C,k}^{\text{dl}H} \mathbf{E}_i
    + \mathbf{\Omega}_i^{\text{dl}}
    \end{array} \right) \right) \nonumber \\
    & -\frac{n_{E,i}}{\ln 2} - \log_2\det\left( \mathbf{\Omega}_i^{\text{dl}} \right), \,\, i\in\mathcal{N}_E. \label{eq:constraint-rewritten-fronthaul-dl-2}
\end{align}
\end{subequations}
The above constraints are equivalent to (\ref{eq:problem-original-fronthaul-dl}) if
\begin{subequations} \label{eq:auxiliary-variable-fronthaul-dl}
\begin{align}
    \alpha^{\text{dl}} & = \frac{\sqrt{\tau_F^{\text{dl}}}}{\tau_E^{\text{dl}}}, \text{ and } \label{eq:auxiliary-variable-fronthaul-dl-1} \\
    \mathbf{\Sigma}_i^{\text{dl}} &=  \sum_{k\in\mathcal{N}_U}  \mathbf{E}_i^H \tilde{\mathbf{Q}}_{C,k}^{\text{dl}}\tilde{\mathbf{Q}}_{C,k}^{\text{dl}H} \mathbf{E}_i
    + \mathbf{\Omega}_i^{\text{dl}}. \label{eq:auxiliary-variable-fronthaul-dl-2}
\end{align}
\end{subequations}

Lastly, using \cite[Cor. 1]{Shen:TN19}, we replace the remaining non-convex constraints (\ref{eq:problem-original-rate-edge-ul-MEC})-(\ref{eq:problem-original-rate-edge-dl-MCC}) with the following stricter constraints:
\begin{subequations} \label{eq:constraint-rewritten-rate-CRAN-ul}
\begin{align}
    \frac{R_{E,k}^{\text{ul}}}{W^{\text{ul}}} & \leq \phi\left( \!\!\! \begin{array}{c} \Gamma_{E,k}^{\text{ul}}\boldsymbol{,} \,\, \boldsymbol{\Theta}_{E,k}^{\text{ul}}\boldsymbol{,} \,\,\tilde{p}_{E,k}^{\text{ul}}\mathbf{h}_{i_k,k}^{\text{ul}}\boldsymbol{,} \\ \begin{array}{c} \sigma_{z,\text{ul}}^2\mathbf{I} + \sum_{l\in\mathcal{N}_U\setminus\{k\}} p_{E,l}^{\text{ul}}\mathbf{h}_{i_k,l}^{\text{ul}}\mathbf{h}_{i_k,l}^{\text{ul}H} \\ + \sum_{l\in\mathcal{N}_U} p_{C,l}^{\text{ul}}\mathbf{h}_{i_k,l}^{\text{ul}}\mathbf{h}_{i_k,l}^{\text{ul}H} \end{array} \end{array}  \!\!\! \right), \label{eq:constraint-rewritten-rate-edge-ul-1} \\
    \frac{R_{C,k}^{\text{ul}}}{W^{\text{ul}}} & \leq \phi\left( \!\!\! \begin{array}{c} \Gamma_{C,k}^{\text{ul}}\boldsymbol{,} \,\, \boldsymbol{\Theta}_{C,k}^{\text{ul}}\boldsymbol{,}\,\, \tilde{p}_{C,k}^{\text{ul}} \mathbf{h}_k^{\text{ul}}\boldsymbol{,}\,\, \\ \begin{array}{c} \sigma_{z,\text{ul}}^2\mathbf{I} + \bar{\boldsymbol{\Omega}}^{\text{ul}} + \sum_{l\in\mathcal{N}_U} p_{E,l}^{\text{ul}}\tilde{\mathbf{h}}_l^{\text{ul}} \tilde{\mathbf{h}}_l^{\text{ul}H} \\ + \sum_{l\in\mathcal{N}_U\setminus\{k\}} p_{C,l}^{\text{ul}}\mathbf{h}_l^{\text{ul}} \mathbf{h}_l^{\text{ul}H} \end{array} \end{array} \!\!\! \right), \\
    \frac{R_{E,k}^{\text{dl}}}{W^{\text{dl}}}& \leq \phi\left( \begin{array}{c} \boldsymbol{\Gamma}_{E,k}^{\text{dl}}\boldsymbol{,}\,\, \boldsymbol{\Theta}_{E,k}^{\text{dl}}\boldsymbol{,}\,\, \mathbf{h}_{k,i_k}^{\text{dl}H}\tilde{\mathbf{Q}}_{E,k}^{\text{dl}}\boldsymbol{,}\,\,  \\ \begin{array}{c} \sigma_{z,\text{dl}}^2 + \mathbf{h}_k^{\text{dl}H}\bar{\mathbf{\Omega}}^{\text{dl}}\mathbf{h}_k^{\text{dl}} + \\ \sum_{l\in\mathcal{N}_U\setminus\{k\}} \mathbf{h}_{k,i_l}^{\text{dl}H}\mathbf{Q}_{E,l}\mathbf{h}_{k,i_l}^{\text{dl}} \\ + \sum_{l\in\mathcal{N}_U} \mathbf{h}_k^{\text{dl}H}\mathbf{Q}_{C,l}\mathbf{h}_k^{\text{dl}}  \end{array}  \end{array}  \right), \text{ and } \\
    \frac{R_{C,k}^{\text{dl}}}{W^{\text{dl}}} & \leq \phi\left( \begin{array}{c} \boldsymbol{\Gamma}_{C,k}^{\text{dl}}\boldsymbol{,}\,\, \boldsymbol{\Theta}_{C,k}^{\text{dl}}\boldsymbol{,}\,\, \mathbf{h}_k^{\text{dl}H}\tilde{\mathbf{Q}}_{C,k}^{\text{dl}}\boldsymbol{,}\,\, \\ \begin{array}{c} \sigma_{z,\text{dl}}^2 + \mathbf{h}_k^{\text{dl}H}\bar{\mathbf{\Omega}}^{\text{dl}}\mathbf{h}_k^{\text{dl}} + \\ \sum_{l\in\mathcal{N}_U} \mathbf{h}_{k,i_l}^{\text{dl}H}\mathbf{Q}_{E,l}\mathbf{h}_{k,i_l}^{\text{dl}} \\ + \sum_{l\in\mathcal{N}_U\setminus\{k\}} \mathbf{h}_k^{\text{dl}H}\mathbf{Q}_{C,l}\mathbf{h}_k^{\text{dl}}  \end{array} \end{array}   \right), \label{eq:constraint-rewritten-rate-edge-ul-4}
\end{align}
\end{subequations}
for $k\in\mathcal{N}_U$.
The above constraints are equivalent to (\ref{eq:problem-original-rate-edge-ul-MEC})-(\ref{eq:problem-original-rate-edge-dl-MCC}) if the variables $\boldsymbol{\Gamma}\triangleq\{ \Gamma_{E,k}^{\text{ul}}, \Gamma_{C,k}^{\text{ul}}, \boldsymbol{\Gamma}_{E,k}^{\text{dl}}, \boldsymbol{\Gamma}_{C,k}^{\text{dl}} \}_{k\in\mathcal{N}_U}$ and $\boldsymbol{\Theta} \triangleq \{ \boldsymbol{\Theta}_{E,k}^{\text{ul}}, \boldsymbol{\Theta}_{C,k}^{\text{ul}}, \boldsymbol{\Theta}_{E,k}^{\text{dl}}, \boldsymbol{\Theta}_{C,k}^{\text{dl}} \}_{k\in\mathcal{N}_U}$ are given as (\ref{eq:update-auxiliary-proposed}) at the bottom of p. 11.

\begin{figure*}[!b]
\hrulefill
\begin{subequations} \label{eq:update-auxiliary-proposed}
\begin{align}
    \Gamma_{E,k}^{\text{ul}} &= p_{E,k}^{\text{ul}} \mathbf{h}_{i_k,k}^{\text{ul}H} \left(  \sigma_{z,\text{ul}}^2\mathbf{I} + \sum_{l\in\mathcal{N}_U\setminus\{k\}} p_{E,l}^{\text{ul}} \mathbf{h}_{i_k,l}^{\text{ul}} \mathbf{h}_{i_k,l}^{\text{ul}H}  + \sum_{l\in\mathcal{N}_U} p_{C,l}^{\text{ul}} \mathbf{h}_{i_k,l}^{\text{ul}} \mathbf{h}_{i_k,l}^{\text{ul}H}  \right)^{-1} \mathbf{h}_{i_k,k}^{\text{ul}}, \label{eq:update-auxiliary-proposed-1} \\
    \boldsymbol{\Theta}_{E,k}^{\text{ul}} &= \tilde{p}_{E,k}^{\text{ul}} \left( \sigma_{z,\text{ul}}^2\mathbf{I} + \sum_{l\in\mathcal{N}_U} p_{E,l}^{\text{ul}} \mathbf{h}_{i_k,l}^{\text{ul}} \mathbf{h}_{i_k,l}^{\text{ul}H} + \sum_{l\in\mathcal{N}_U} p_{C,l}^{\text{ul}} \mathbf{h}_{i_k,l}^{\text{ul}} \mathbf{h}_{i_k,l}^{\text{ul}H} \right)^{-1} \mathbf{h}_{i_k,k}^{\text{ul}}, \label{eq:update-auxiliary-proposed-2} \\
    \Gamma_{C,k}^{\text{ul}} &= p_{C,k}^{\text{ul}} \mathbf{h}_k^{\text{ul}H} \left( \sigma_{z,\text{ul}}^2\mathbf{I} + \bar{\mathbf{\Omega}}^{\text{ul}} + \sum_{l\in\mathcal{N}_U} p_{E,l}^{\text{ul}} \tilde{\mathbf{h}}_l^{\text{ul}}\tilde{\mathbf{h}}_l^{\text{ul}H} + \sum_{l\in\mathcal{N}_U\setminus\{k\}} p_{C,l}^{\text{ul}} \mathbf{h}_l^{\text{ul}} \mathbf{h}_l^{\text{ul}H} \right)^{-1} \mathbf{h}_k^{\text{ul}}, \label{eq:update-auxiliary-proposed-3} \\
    \boldsymbol{\Theta}_{C,k}^{\text{ul}} &= \tilde{p}_{C,k}^{\text{ul}} \left( \sigma_{z,\text{ul}}^2\mathbf{I} + \bar{\mathbf{\Omega}}^{\text{ul}} + \sum_{l\in\mathcal{N}_U} p_{E,l}^{\text{ul}} \tilde{\mathbf{h}}_l^{\text{ul}}\tilde{\mathbf{h}}_l^{\text{ul}H} + \sum_{l\in\mathcal{N}_U} p_{C,l}^{\text{ul}} \mathbf{h}_l^{\text{ul}} \mathbf{h}_l^{\text{ul}H} \right)^{-1} \mathbf{h}_k^{\text{ul}}, \label{eq:update-auxiliary-proposed-4} \\
    \boldsymbol{\Gamma}_{E,k}^{\text{dl}} &= \tilde{\mathbf{Q}}_{E,k}^{\text{dl}H}\mathbf{h}_{k,i_k}^{\text{dl}} \left(  \sigma_{z,\text{dl}}^2 + \mathbf{h}_k^{\text{dl}H}\bar{\mathbf{\Omega}}^{\text{dl}}\mathbf{h}_k^{\text{dl}} + \sum_{l\in\mathcal{N}_U\setminus\{k\}} \mathbf{h}_{k,i_l}^{\text{dl}H}\mathbf{Q}_{E,l}\mathbf{h}_{k,i_l}^{\text{dl}} + \sum_{l\in\mathcal{N}_U} \mathbf{h}_k^{\text{dl}H}\mathbf{Q}_{C,l}\mathbf{h}_k^{\text{dl}} \right)^{-1} \mathbf{h}_{k,i_k}^{\text{dl}H} \tilde{\mathbf{Q}}_{E,k}^{\text{dl}}, \label{eq:update-auxiliary-proposed-5} \\
    \boldsymbol{\Theta}_{E,k}^{\text{dl}} &= \left( \sigma_{z,\text{dl}}^2 + \mathbf{h}_k^{\text{dl}H}\bar{\mathbf{\Omega}}^{\text{dl}}\mathbf{h}_k^{\text{dl}} + \sum_{l\in\mathcal{N}_U} \mathbf{h}_{k,i_l}^{\text{dl}H}\mathbf{Q}_{E,l}\mathbf{h}_{k,i_l}^{\text{dl}} + \sum_{l\in\mathcal{N}_U} \mathbf{h}_k^{\text{dl}H}\mathbf{Q}_{C,l}\mathbf{h}_k^{\text{dl}} \right)^{-1} \mathbf{h}_{k,i_k}^{\text{dl}H} \tilde{\mathbf{Q}}_{E,k}^{\text{dl}}, \label{eq:update-auxiliary-proposed-6} \\
    \boldsymbol{\Gamma}_{C,k}^{\text{dl}} &= \tilde{\mathbf{Q}}_{C,k}^{\text{dl}H} \mathbf{h}_k^{\text{dl}} \left(  \sigma_{z,\text{dl}}^2 + \mathbf{h}_k^{\text{dl}H}\bar{\mathbf{\Omega}}^{\text{dl}}\mathbf{h}_k^{\text{dl}} + \sum_{l\in\mathcal{N}_U} \mathbf{h}_{k,i_l}^{\text{dl}H}\mathbf{Q}_{E,l}\mathbf{h}_{k,i_l}^{\text{dl}} + \sum_{l\in\mathcal{N}_U\setminus\{k\}} \mathbf{h}_k^{\text{dl}H}\mathbf{Q}_{C,l}\mathbf{h}_k^{\text{dl}}  \right)^{-1} \mathbf{h}_k^{\text{dl}H} \tilde{\mathbf{Q}}_{C,k}^{\text{dl}}, \label{eq:update-auxiliary-proposed-7} \\
    \boldsymbol{\Theta}_{C,k}^{\text{dl}} &= \left( \sigma_{z,\text{dl}}^2 + \mathbf{h}_k^{\text{dl}H}\bar{\mathbf{\Omega}}^{\text{dl}}\mathbf{h}_k^{\text{dl}} + \sum_{l\in\mathcal{N}_U} \mathbf{h}_{k,i_l}^{\text{dl}H}\mathbf{Q}_{E,l}\mathbf{h}_{k,i_l}^{\text{dl}} + \sum_{l\in\mathcal{N}_U} \mathbf{h}_k^{\text{dl}H}\mathbf{Q}_{C,l}\mathbf{h}_k^{\text{dl}}  \right)^{-1} \mathbf{h}_k^{\text{dl}H} \tilde{\mathbf{Q}}_{C,k}^{\text{dl}}. \label{eq:update-auxiliary-proposed-8}
\end{align}
\end{subequations}
\end{figure*}


\begin{algorithm}
\caption{Alternating optimization algorithm that tackles problem (\ref{eq:problem-original})}

\textbf{1.} Initialize $\{\mathbf{p}, \mathbf{c}, \mathbf{Q}, \mathbf{\Omega}, \boldsymbol{\tau}, \mathbf{R}\}$ as arbitrary matrices/values that satisfy the constraints (\ref{eq:problem-original-latency-edge-ul-MEC})-(\ref{eq:problem-original-rate-edge-dl-MCC}), and set $t\leftarrow 1$.

\textbf{2.} Calculate the total latency $\tau_T$ in (\ref{eq:total-latency}) with the initialized $\{\mathbf{p}, \mathbf{c}, \mathbf{Q}, \mathbf{\Omega}, \boldsymbol{\tau}, \mathbf{R}\}$, and set $\tau_T^{(0)}\leftarrow \tau_T$.

\textbf{3.} Set $\{\boldsymbol{\lambda}, \boldsymbol{\alpha}, \boldsymbol{\Sigma}, \boldsymbol{\Gamma}, \boldsymbol{\Theta}\}$ according to (\ref{eq:update-lambda-TDMA}), (\ref{eq:update-lambda-latency-edge-proposed}), (\ref{eq:auxiliary-variable-fronthaul-ul}), (\ref{eq:auxiliary-variable-fronthaul-dl}) and (\ref{eq:update-auxiliary-proposed}).

\textbf{4.} Update $\{\mathbf{p}, \mathbf{c}, \mathbf{Q}, \mathbf{\Omega}, \boldsymbol{\gamma}, \boldsymbol{\tau}, \mathbf{R}\}$ as a solution of the convex problem which is obtained from (\ref{eq:problem-original}) by replacing the constraints (\ref{eq:problem-original-latency-edge-ul-MEC})-(\ref{eq:problem-original-rate-edge-dl-MCC}) with (\ref{eq:rewrite-latency-constraint-TDMA-3}), (\ref{eq:rewrite-latenccy-constraint-TDMA-4}), (\ref{eq:consraints-rewritten-latency-proposed}), (\ref{eq:constraint-rewritten-fronthaul-ul}), (\ref{eq:constraint-rewritten-fronthaul-dl}) and (\ref{eq:constraint-rewritten-rate-CRAN-ul}), and fixing the variables $\{\boldsymbol{\lambda}, \boldsymbol{\alpha}, \boldsymbol{\Sigma}, \boldsymbol{\Gamma}, \boldsymbol{\Theta}\}$.

\textbf{5.} Calculate the total latency $\tau_T$ in (\ref{eq:total-latency}) with the updated $\{\mathbf{p}, \mathbf{c}, \mathbf{Q}, \mathbf{\Omega}, \boldsymbol{\gamma}, \boldsymbol{\tau}, \mathbf{R}\}$, and set $\tau_T^{(t)}\leftarrow \tau_T$.

\textbf{6.} Stop if $|\tau_T^{(t)} - \tau_T^{(t-1)}|\leq\delta$ or $t>t_{\max}$. Otherwise, set $t\leftarrow t+1$ and go back to Step 3.

\end{algorithm}

Based on the discussed inequalities (\ref{eq:rewrite-latency-constraint-TDMA-3}), (\ref{eq:rewrite-latenccy-constraint-TDMA-4}), (\ref{eq:consraints-rewritten-latency-proposed}), (\ref{eq:constraint-rewritten-fronthaul-ul}), (\ref{eq:constraint-rewritten-fronthaul-dl}), and (\ref{eq:constraint-rewritten-rate-CRAN-ul}) that restate the non-convex constraints (\ref{eq:problem-original-latency-edge-ul-MEC})-(\ref{eq:problem-original-rate-edge-dl-MCC}) of problem (\ref{eq:problem-original}), we propose an iterative algorithm that alternately optimizes $\{\mathbf{p}, \mathbf{c}, \mathbf{Q}, \mathbf{\Omega}, \boldsymbol{\tau}, \mathbf{R}\}$ and $\{\boldsymbol{\lambda}, \boldsymbol{\gamma}, \boldsymbol{\alpha}, \boldsymbol{\Sigma}, \boldsymbol{\Gamma}, \boldsymbol{\Theta}\}$. When optimizing $\{\mathbf{p}, \mathbf{c}, \mathbf{Q}, \mathbf{\Omega}, \boldsymbol{\tau}, \mathbf{R}\}$, we tackle the convex problem which is obtained from (\ref{eq:problem-original}) by replacing the constraints (\ref{eq:problem-original-latency-edge-ul-MEC})-(\ref{eq:problem-original-rate-edge-dl-MCC}) with (\ref{eq:rewrite-latency-constraint-TDMA-3}), (\ref{eq:rewrite-latenccy-constraint-TDMA-4}), (\ref{eq:consraints-rewritten-latency-proposed}), (\ref{eq:constraint-rewritten-fronthaul-ul}), (\ref{eq:constraint-rewritten-fronthaul-dl}) and (\ref{eq:constraint-rewritten-rate-CRAN-ul}), and fixing the variables $\{\boldsymbol{\lambda}, \boldsymbol{\gamma}, \boldsymbol{\alpha}, \boldsymbol{\Sigma}, \boldsymbol{\Gamma}, \boldsymbol{\Theta}\}$. For fixed $\{\mathbf{p}, \mathbf{c}, \mathbf{Q}, \mathbf{\Omega}, \boldsymbol{\tau}, \mathbf{R}\}$, the optimal variables $\{\boldsymbol{\lambda}, \boldsymbol{\gamma}, \boldsymbol{\alpha}, \boldsymbol{\Sigma}, \boldsymbol{\Gamma}, \boldsymbol{\Theta}\}$ are obtained as (\ref{eq:update-lambda-TDMA}), (\ref{eq:update-lambda-latency-edge-proposed}), (\ref{eq:auxiliary-variable-fronthaul-ul}), (\ref{eq:auxiliary-variable-fronthaul-dl}) and (\ref{eq:update-auxiliary-proposed}). The detailed algorithm is described in Algorithm 3. 
The solution obtained by Algorithm 3 is a locally optimal solution due to the non-convexity of the problem (\ref{eq:problem-original}).
In Sec. \ref{sec:numerical}, we initialize $\{\mathbf{p}, \mathbf{c}\}$ as $p_{E,k}^{\text{ul}}\leftarrow P^{\text{ul}}$, $p_{C,k}^{\text{ul}}\leftarrow P^{\text{ul}}$ and $c_k\leftarrow 1/2$ for $k\in\mathcal{N}_U$.
To initialize the covariance matrices of downlink signals $\mathbf{Q}$ and quantization noise signals $\boldsymbol{\Omega}$, we first set
\begin{subequations} \label{eq:init-covariance-CRAN}
\begin{align}
    &\mathbf{Q}_{E,k} \leftarrow \mathbf{V}_{E,k}\mathbf{V}_{E,k}^H, \, k\in\mathcal{N}_{U,i}, i\in\mathcal{N}_E, \label{eq:init-QE-CRAN} \\
    &\mathbf{Q}_{C,k} \leftarrow \mathbf{V}_{C,k}\mathbf{V}_{C,k}^H, \, k\in\mathcal{N}_{U}, \label{eq:init-QC-CRAN}\\
    &\boldsymbol{\Omega}_i \leftarrow \mathbf{V}_{\Omega,i}\mathbf{V}_{\Omega,i}^H, \, i\in\mathcal{N}_E, \label{eq:init-Omega-CRAN}
\end{align}
\end{subequations}
where the elements of $\mathbf{V}_{E,k}\in\mathbb{C}^{n_{E,i}\times n_{E,i}}$, $\mathbf{V}_{C,k}\in\mathbb{C}^{n_{E}\times n_{E}}$ and $\mathbf{V}_{\Omega,k}\in\mathbb{C}^{n_{E,i}\times n_{E,i}}$ follow $\mathcal{CN}(0,1)$.
The covariance matrices obtained in (\ref{eq:init-covariance-CRAN}) may not satisfy the power constraints (\ref{eq:power-constraint-dl-rewritten}). To resolve this issue, we repeatedly multiply a scalar $\eta < 1$ to the matrices $\mathbf{Q}$ and $\boldsymbol{\Omega}$ until the constraints (\ref{eq:power-constraint-dl-rewritten}) are satisfied. In the simulation, we set $\eta=1/2$.
Once the variables $\{\mathbf{p}, \mathbf{c}, \mathbf{Q}, \boldsymbol{\Omega}\}$ are fixed, the rate variables $\mathbf{R}$ can be computed using (\ref{eq:achievable-rate-MEC-ul}), (\ref{eq:achievable-rate-MCC-ul}) and (\ref{eq:achievable-rate-dl}), and the latency variables $\boldsymbol{\tau}$ are initialized as (\ref{eq:fronthaul-latency-uplink}), (\ref{eq:latency-edge-MEC-uplink}), (\ref{eq:fronthaul-latency-dl}), and (\ref{eq:latency-edge-MEC-downlink}).

As discussed in Sec. \ref{sec:D-RAN}, the complexity of Algorithm 3 is given by the number of iterations multiplied by the complexity of solving the convex problem at Step 4. 
The complexity of the latter is upper bounded by $\mathcal{O}( n(n^3+M) \log(1/\epsilon) )$ \cite[p. 4]{BTal-Nemirovski}, where the numbers $n$ and $M$ equal $n = N_{U}(4\tilde{n}_{E}^{2}(N_{E}^{2}+1)+10)+N_{E}(8\tilde{n}_{E}^{2}+2)+6$ and $M = \left(8\tilde{n}_{E}^{2}N_{U}+D_{\tilde{n}_{E}}\right)N_{E}+ 4N_{U}N_{E}^{2}\tilde{n}_{E}^{2}\left(8N_{E}\tilde{n}_{E}+3N_{U}\right)+50N_{U}+5N_{E}\tilde{n}_{E}$, respectively.
Here $D_L$ is defined as the number of arithmetic operations needed to calculate the determinant of an $L\times L$ matrix, which is given as $D_L = \mathcal{O}(L^3)$ with Gaussian elimination \cite[p. 1]{Rote}.
We discuss the convergence rate of Algorithm 3 in Sec. \ref{sec:numerical}.

\section{Numerical Results} \label{sec:numerical}

In this section, we validate via numerical results the performance gain of the proposed C-RAN architecture as compared to the D-RAN reference system. We assume that the locations of $N_U$ UEs and $N_E$ ENs are independently and uniformly sampled from a square area with side length of 500 m. We impose the minimum separation of 10 m between any pair of UE and EN. We consider a path-loss model $\rho_0 (d / d_0)^{-\eta}$ \cite{Jeon-et-al:CL19, Kim-Park:WCL20}, where $\rho_0$ is the path-loss at a reference distance $d_0$, $d$ denotes the distance between the transmitting and receiving nodes, and $\eta$ is the path-loss exponent. We set $d_0=30$ m, $\rho_0=10$ dB and $\eta=3$, and assume independent Rayleigh small-scale fading channel model for all the channel coefficients. We consider a symmetric system between uplink and downlink with $\text{SNR}^{\text{ul}}_{\max} = \text{SNR}^{\text{dl}}_{\max} = \text{SNR}_{\max}$, $W^{\text{ul}}=W^{\text{dl}} = W$, and $C_F^{\text{ul}} = C_F^{\text{dl}} = C_F$.
The computation capabilities of CP and ENs are set to $F_C=10^{11}$ \cite{Ashuwaili:TSIPN17} and $F_{E,i}\in\{1.0, 2.5\}\times 10^{10}$ \cite{Ashuwaili:WCL17, Chen-Hao:JSAC18}, respectively, unless stated otherwise. We also assume that there are $b_{I,k}=b_{O,k}=10^6$ input and output bits for each UE and that the task of each UE $k$ requires 
$V_k=700$ CPU cycles per input bit \cite{Ren-et-al:TVT19}.
To solve the convex problems at Step 4 of Algorithms 1, 2 and 3, CVX software \cite{CVX} with SDPT3 solver \cite{SDPT3} is adopted.
Without claim of optimality, we associate each UE $k$ with the closest EN, so that $i_k$ is set to
\begin{align}
    i_k = \arg \min_{i\in\mathcal{N}_E} \text{dist}_{i,k},
\end{align}
with $\text{dist}_{i,k}$ represents the geographical distance between UE $k$ and EN $i$.

\subsection{Convergence of the Proposed Algorithm} \label{sub:convergence}

\begin{figure}
$\!\!\!$%
\begin{minipage}[t]{1.0\columnwidth}%
\centering\includegraphics[width=9cm,height=7cm,keepaspectratio]{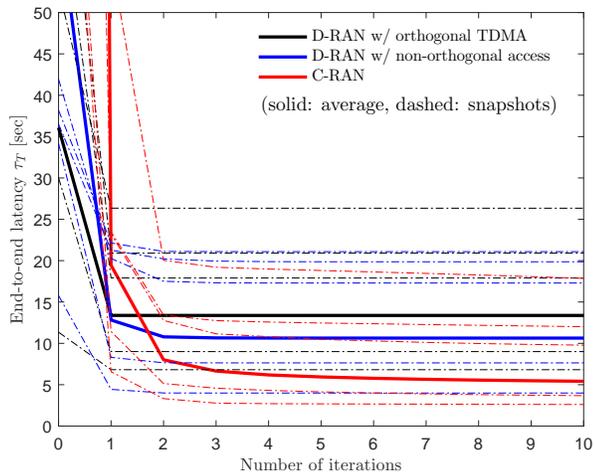}

~~~\centering{\footnotesize{}(a) $\text{SNR}_{\max}=0$ dB}{\footnotesize \par}%
\end{minipage}~~~~%
\\
\begin{minipage}[t]{1.0\columnwidth}%
\centering\includegraphics[width=9cm,height=7cm,keepaspectratio]{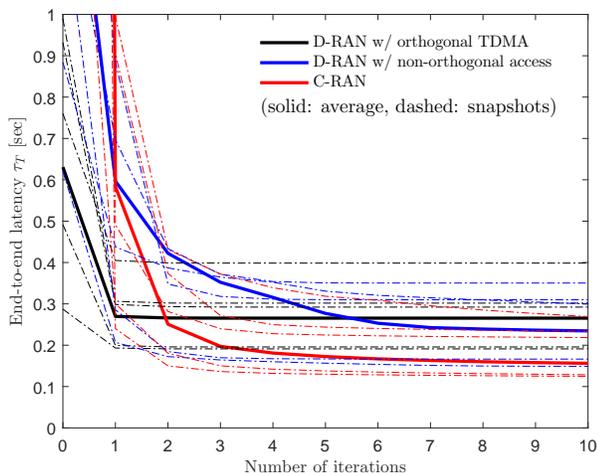}

~~\centering{\footnotesize{}(b) $\text{SNR}_{\max}=20$ dB}{\footnotesize \par}%
\end{minipage}

\caption{{\footnotesize{}\label{fig:graph-tauT-vs-numItr}End-to-end latency $\tau_T$ versus
the number of iterations ($N_U=4$, $N_E=2$, $n_{E,i}=2$, $W=20$ MHz, $C_F=1$ Gbps, $F_{E,i}=10^{10}$ and $\text{SNR}_{\max}\in\{0,20\}$ dB).}}
\end{figure}

The convergence rate of FP is analyzed in \cite{Shen-Yu:TSP18} with a focus on single-ratio problems, and reference \cite{Shen:TN19} discusses the convergence rate of matrix FP via numerical examples. 
Similar to \cite{Shen:TN19}, we provide numerical evidence of the fast convergence of the proposed algorithms in Fig.  \ref{fig:graph-tauT-vs-numItr}. In the figure, we plot the end-to-end latency $\tau_T$ of D-RAN and C-RAN schemes versus the number of iterations for $N_U=4$, $N_E=2$, $n_{E,i}=2$, $W=20$ MHz, $C_F=1$ Gbps, $F_{E,i}=10^{10}$ and $\text{SNR}_{\max}\in\{0,20\}$ dB. 
We plot both the snapshots and average latency, where the latter is averaged over 100 channel samples.
The figure shows that, regardless of the SNR, the proposed algorithms converge reliably within a few iterations. 
We leave the analysis of the convergence rate of the proposed algorithms for future work.
Throughout the following experiments, we set the threshold value for convergence as $\delta=10^{-4}$ and limit the maximum number of iterations to $t_{\max}=30$.

\subsection{Performance Gains of the C-RAN Architecture} \label{sub:D-RAN-vs-C-RAN}

\begin{figure}
\centering\includegraphics[width=9cm,height=7cm,keepaspectratio]{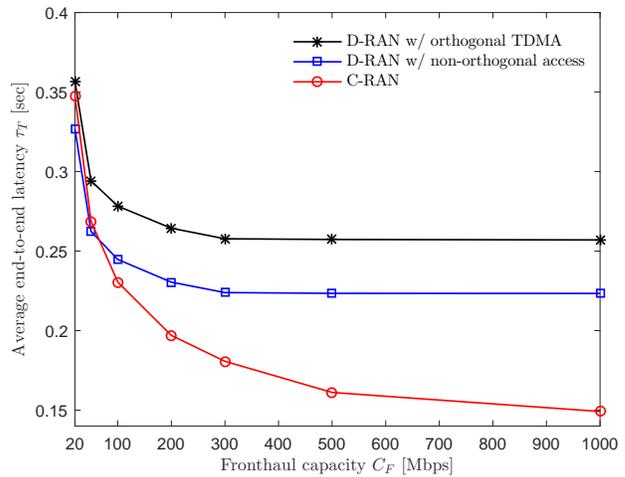}\caption{{\footnotesize{}\label{fig:graph-DRANvsCRAN-tauT-vs-CF}Average end-to-end latency $\tau_T$ versus
the fronthaul capacity $C_F$ ($N_U=4$, $N_E=2$, $n_{E,i}=2$, $W=20$ MHz, $F_{E,i}=10^{10}$ and $\text{SNR}_{\max}=20$ dB).}}
\end{figure}

In this subsection, we investigate the performance gains of the C-RAN architecture introduced in Sec. \ref{sec:C-RAN} for collaborative cloud and edge mobile computing as compared to benchmark D-RAN systems described in Sec. \ref{sec:D-RAN}. To this end, in Fig. \ref{fig:graph-DRANvsCRAN-tauT-vs-CF}, we plot the average end-to-end latency $\tau_T$ versus the fronthaul capacity $C_F$ for $N_U=4$, $N_E=2$, $n_{E,i}=2$, $W=20$ MHz, $F_{E,i}=10^{10}$ and $\text{SNR}_{\max}=20$ dB.
The figure shows that deploying C-RAN architecture is not advantageous when the fronthaul capacity $C_F$ is small due to the large latency caused by the fronthaul transmission. However, as $C_F$ increases, the C-RAN scheme significantly outperforms the benchmark D-RAN schemes, since it enables more effective interference management by means of centralized encoding and decoding at CP.

\begin{figure}
\centering\includegraphics[width=9cm,height=7cm,keepaspectratio]{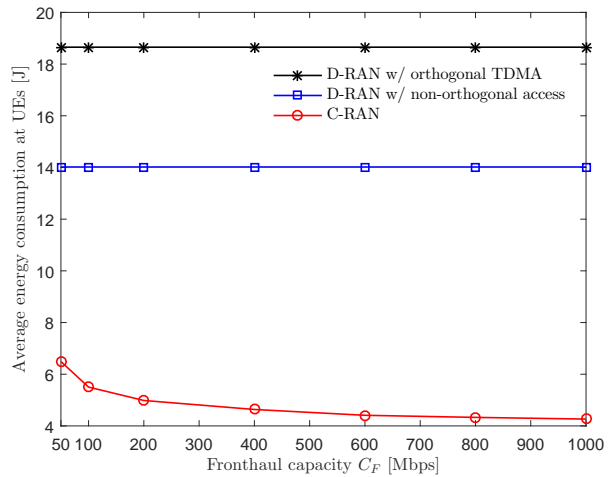}\caption{{\footnotesize{}\label{fig:graph-DRANvsCRAN-energy-vs-CF}Average energy consumption at UEs versus
the fronthaul capacity $C_F$ ($N_U=4$, $N_E=2$, $n_{E,i}=2$, $W=20$ MHz, $F_{E,i}=10^{10}$ and $\text{SNR}_{\max}=20$ dB).}}
\end{figure}

In Fig. \ref{fig:graph-DRANvsCRAN-energy-vs-CF}, we examine the energy consumption at UEs under the same set-up considered in Fig. \ref{fig:graph-DRANvsCRAN-tauT-vs-CF}. We calculate the energy consumption at UE $k$ as $E_k = E_k^{\text{ul}} + E_k^{\text{dl}}$, where the uplink and downlink energy expenditures are defined as $E_k^{\text{ul}} = \tau_{E,k}^{\text{ul}} \tilde{p}_k^{\text{ul}}$ and $E_k^{\text{dl}} = \tau_{E,k}^{\text{dl}} d_k^{\text{dl}}$, respectively. Here, $d_k^{\text{dl}}$ indicates the mobile receiving energy expenditure per second in downlink, and is set to $d_k = 0.625$ J/s as in \cite{Ashuwaili:WCL17}. The uplink transmit power $\tilde{p}_k^{\text{ul}}$ of UE $k$ is respectively given as $\tilde{p}_k^{\text{ul}} = p_k^{\text{ul}}$ and $\tilde{p}_k^{\text{ul}} = p_{E,k}^{\text{ul}} + p_{C,k}^{\text{ul}}$ for the D-RAN and C-RAN systems.
Unlike D-RAN, the energy consumption of UEs with C-RAN decreases with $C_F$. This is because the ENs and CP can exchange quantized baseband signals of better resolution for larger $C_F$, and hence the latency on edge links becomes lower.

\begin{figure}
\centering\includegraphics[width=9cm,height=7cm,keepaspectratio]{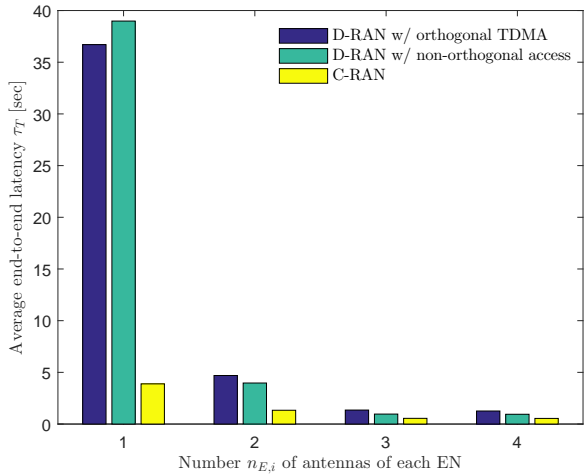}\caption{{\footnotesize{}\label{fig:graph-DRANvsCRAN-tauT-vs-nEi}Average end-to-end latency $\tau_T$ versus
the number $n_{E,i}$ of antennas of each EN ($N_U=3$, $N_E=2$, $W=20$ MHz, $F_{E,i}=10^{10}$, $C_F=3$ Gbps and $\text{SNR}_{\max}=5$ dB).}}
\end{figure}

Fig. \ref{fig:graph-DRANvsCRAN-tauT-vs-nEi} plots the average end-to-end latency $\tau_T$ with respect to the number $n_{E,i}$ of antennas of each EN for $N_U=3$, $N_E=2$, $W=20$ MHz, $F_{E,i}=10^{10}$, $C_F=3$ Gbps and $\text{SNR}_{\max}=5$ dB.
Comparing the performance of D-RAN with different access techniques, we see that TDMA shows a lower latency than non-orthogonal access when the ENs use a small number of antennas. However, when the ENs are equipped with sufficiently many antennas, the non-orthogonal scheme outperforms the TDMA scheme, since the co-channel interference signals can be suppressed by local array processing at the ENs. In this case, each EN can suppress interference signals only with local processing, and hence C-RAN does not provide performance benefits, while significant gains are observed for lower values of $n_{E,i}$.

\begin{figure}
\centering\includegraphics[width=9cm,height=7cm,keepaspectratio]{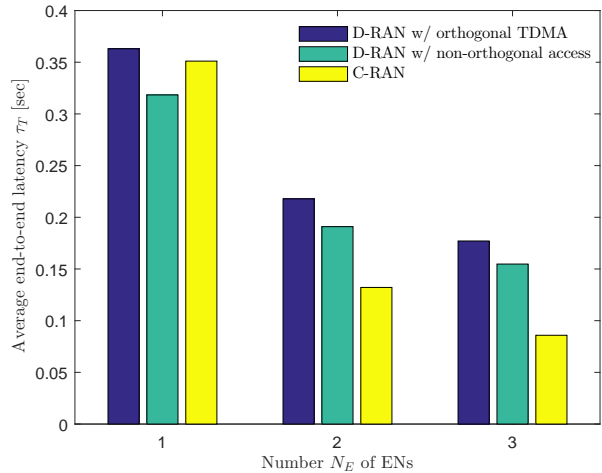}\caption{{\footnotesize{}\label{fig:graph-DRANvsCRAN-tauT-vs-NE}Average end-to-end latency $\tau_T$ versus
the number $N_E$ of ENs ($N_U=8$, $n_{E,i}=2$, $W=50$ MHz, $F_{E,i}=2.5\times 10^{10}$, $C_F=2$ Gbps and $\text{SNR}_{\max}=20$ dB).}}
\end{figure}

In Fig. \ref{fig:graph-DRANvsCRAN-tauT-vs-NE}, we plot the average end-to-end latency $\tau_T$ versus the number $N_E$ of ENs for $N_U=8$, $n_{E,i}=2$, $W=50$ MHz, $F_{E,i}=2.5\times 10^{10}$, $C_F=2$ Gbps and $\text{SNR}_{\max}=20$ dB.
When the network has a single EN, i.e., $N_E=1$, there is no advantage of deploying the C-RAN architecture in Sec. \ref{sec:C-RAN} compared to D-RAN in Sec. \ref{sec:D-RAN}. This is because the noise signals caused by fronthaul quantization degrade the spectral efficiency for both uplink and downlink.
However, as $N_E$ increases, C-RAN shows significantly improved latency performance than the D-RAN schemes. These gains are achieved by the centralized signal processing at the CP on behalf of the connected ENs, which enables effective interference management.

\subsection{Performance Gains of Collaborative Cloud-Edge Computing} \label{sub:collaborative-MEC-MCC}

In this subsection, we study the performance gains of the collaborative cloud and edge computing system with optimized computational resource allocation as compared to benchmark schemes that rely only on edge computing (i.e., by setting $c_k=1$ for all $k\in\mathcal{N}_U$) or cloud computing (i.e., $c_k=0$ for all $k\in\mathcal{N}_U$). Note that the optimization of these benchmark schemes can be addressed by adopting the proposed algorithm with minor modifications. For reference, we also evaluate the performance of a hybrid strategy that selects between the two benchmark schemes. We adopt the optimized C-RAN architecture in Sec. \ref{sec:C-RAN} for all cases except for edge computing, for which the C-RAN system is not applicable and hence we select D-RAN with non-orthogonal multiple access.

\begin{figure}
\centering\includegraphics[width=9cm,height=7cm,keepaspectratio]{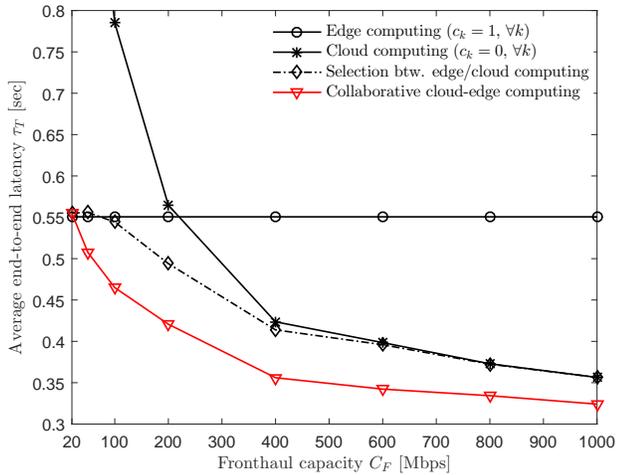}\caption{{\footnotesize{}\label{fig:graph-MECvsMCCvsJoint-tauT-vs-CF}Average end-to-end latency $\tau_T$ versus
the fronthaul capacity $C_F$ ($N_U=4$, $N_E=2$, $n_{E,i}=2$, $W=50$ MHz, $F_{E,i}=2.5\times 10^{10}$ and $\text{SNR}_{\max}=10$ dB).}}
\end{figure}

In Fig. \ref{fig:graph-MECvsMCCvsJoint-tauT-vs-CF}, we plot the average end-to-end latency $\tau_T$ versus the fronthaul capacity $C_F$ for $N_U=4$, $N_E=2$, $n_{E,i}=2$, $W=50$ MHz, $F_{E,i}=2.5\times 10^{10}$ and $\text{SNR}_{\max}=10$ dB.
Since edge computing does not utilize the fronthaul links, its performance is not affected by $C_F$. 
In contrast, the latency of cloud computing scheme decreases as $C_F$ increases. 
While selecting between edge and cloud computing schemes does not yield significant benefits, the proposed collaborative cloud and edge scheme achieves notable gains, particularly in the intermediate regime of $C_F$.

\begin{figure}
\centering\includegraphics[width=9cm,height=7cm,keepaspectratio]{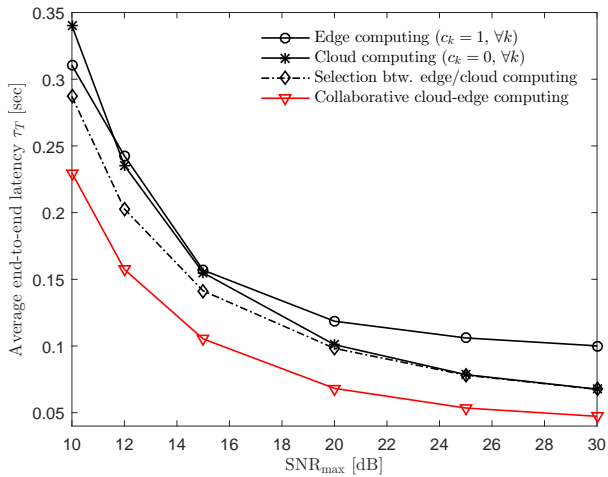}\caption{{\footnotesize{}\label{fig:graph-MECvsMCCvsJoint-tauT-vs-SNR}Average end-to-end latency $\tau_T$ versus
the maximum SNR ($N_U=4$, $N_E=2$, $n_{E,i}=2$, $W=100$ MHz, $F_{E,i}=2.5\times 10^{10}$ and $C_F=250$ Mbps).}}
\end{figure}

In Fig. \ref{fig:graph-MECvsMCCvsJoint-tauT-vs-SNR}, we plot the average end-to-end latency $\tau_T$ versus the maximum SNR for $N_U=4$, $N_E=2$, $n_{E,i}=2$, $W=100$ MHz, $F_{E,i}=2.5\times 10^{10}$ and $C_F=250$ Mbps. The figure shows that, although increased SNR levels are beneficial for all the schemes, the performance of cloud computing is more significantly affected by the SNR than that of edge computing. This is because the edge latency of edge computing is limited by interference, and hence its performance saturates as the SNR increases. The performance of the C-RAN scheme is instead limited by the fronthaul capacity as SNR grows larger.

\begin{figure}
\centering\includegraphics[width=9cm,height=7cm,keepaspectratio]{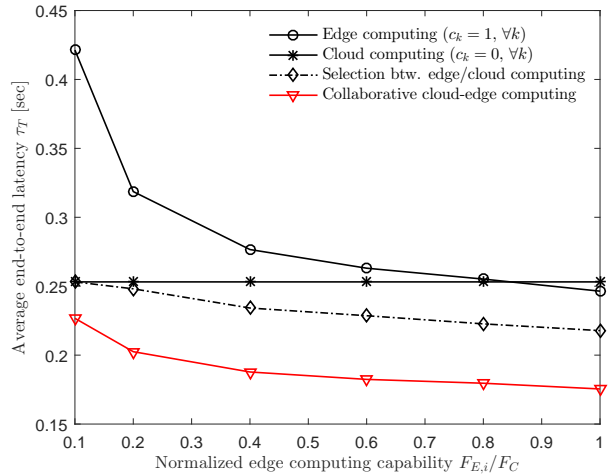}\caption{{\footnotesize{}\label{fig:graph-MECvsMCCvsJoint-tauT-vs-FEi}Average end-to-end latency $\tau_T$ versus
the normalized edge computing capability $F_{E,i}/F_C$ ($N_U=4$, $N_E=2$, $n_{E,i}=2$, $W=100$ MHz, $C_F=500$ Mbps,  $\text{SNR}_{\max}=10$ dB and $F_C=10^{11}$).}}
\end{figure}

Fig. \ref{fig:graph-MECvsMCCvsJoint-tauT-vs-FEi} plots the average end-to-end latency $\tau_T$ by varying the edge computing capability $F_{E,i}$ normalized by $F_C$ for $N_U=4$, $N_E=2$, $n_{E,i}=2$, $W=100$ MHz, $C_F=500$ Mbps,  $\text{SNR}_{\max}=10$ dB and $F_C=10^{11}$.
When $F_{E,i}$ is too small, it is desired to choose $c_k=0$ for all $k\in\mathcal{N}_U$ so that all the tasks are offloaded to the CP. As $F_{E,i}$ increases, offloading some tasks to ENs can improve the performance, and the proposed scheme with optimized task allocation provides a notable gain as compared to all the benchmark schemes.

\begin{figure}
\centering\includegraphics[width=9cm,height=7cm,keepaspectratio]{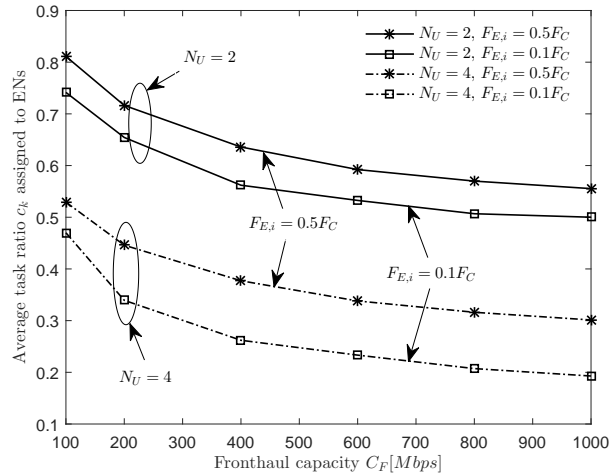}\caption{{\footnotesize{}\label{fig:graph-ck-vs-CF}Average task ratio $c_k$ assigned to ENs versus
the fronthaul capacity $C_F$ ($N_U\in\{2,4\}$, $N_E=2$, $n_{E,i}=1$, $W=100$ MHz and $F_{E,i}\in\{0.1, 0.5\}\times 10^{10}$).}}
\end{figure}

In Fig. \ref{fig:graph-ck-vs-CF}, we plot the average task ratio $c_k$ assigned to ENs versus the fronthaul capacity $C_F$ for $N_U\in\{2,4\}$, $N_E=2$, $n_{E,i}=1$, $W=100$ MHz and $F_{E,i}\in\{0.1, 0.5\}\times 10^{10}$. The task ratio variables are obtained from the proposed algorithm in Sec. \ref{sub:optimization}. We observe from the figure that, as the fronthaul capacity $C_F$ increases, more tasks are assigned to CP due to reduced fronthaul latency. Similarly, as the ENs are equipped with  stronger computing power $F_{E,i}$, they process a larger portion of tasks. Moreover, increasing the number $N_U$ of UEs results in  smaller ratios $c_k$, since the ENs with limited computing power offload more tasks to the CP when $N_U$ is larger.

\section{Conclusions} \label{sec:conclusion}

We have studied the design of collaborative cloud and edge mobile computing within a C-RAN architecture for minimal end-to-end latency. We have tackled the joint design of computational resource allocation and C-RAN signal processing strategies with the goal of minimizing end-to-end latency required for completing the computational tasks of all the participating UEs in the network. To tackle the non-convex optimization problem, we have applied FP and matrix FP. Via extensive numerical results, we have validated the convergence of the proposed optimization algorithms, the performance gain of C-RAN architecture as compared to D-RAN, and the impact of optimized computational resource allocation of collaborative cloud and edge computing. As future work, we mention the extension to collaborative AR \cite{Ashuwaili:WCL17}, heterogeneous C-RAN and mobile computing integrated systems \cite{Peng-et-al:WC, Pham-et-al:Access, Ma-et-al:Access}, the robust design with imperfect CSI \cite{Gharavol-Larsson:TSP13}, and the energy-efficient design \cite{Sardellitti:TSIPN15, Ashuwaili:TSIPN17} for energy-limited mobile UEs.
Also, it would be relevant to verify the effectiveness of the proposed algorithms by deriving a tight lower bound on the optimal latency values.

\end{document}